\documentclass{aa}
\pdfoutput=1
\usepackage[T1]{fontenc}
\usepackage[hyperindex,pdftex]{hyperref}
\usepackage[figure, figure*]{hypcap}
\usepackage{graphicx}
\usepackage{epstopdf}
\usepackage{amssymb}
\usepackage{float}
\usepackage{color}
\usepackage{geometry}
\usepackage{marginnote}
\usepackage{amsmath}
\usepackage{natbib}
\usepackage{bibentry}
\usepackage{array} 
\usepackage[varg]{txfonts}
\usepackage{caption}
\usepackage{bm}
\usepackage{indentfirst}
\usepackage{subcaption}
\usepackage{xcolor}
\usepackage[super]{nth}
\usepackage[labelfont=color]{caption}
\DeclareMathSymbol{\varOmega}{\mathord}{letters}{"0A}
\DeclareMathSymbol{\varPsi}{\mathord}{letters}{"09}
\DeclareMathSymbol{\varPhi}{\mathord}{letters}{"08}
\DeclareMathSymbol{\varGamma}{\mathord}{letters}{"00}
\DeclareMathSymbol{\varPi}{\mathord}{letters}{"05}
\DeclareMathSymbol{\varLambda}{\mathord}{letters}{"03}
\DeclareCaptionFont{red}{\color{red}}

\graphicspath{{./Plots_only/}}
\begin{document}

\title{Streaming Instability of Multiple Particle Species \break in Protoplanetary Disks}
\author{Noemi Schaffer
\and Chao-Chin Yang\thanks{\textit{Present address}: Department of Physics and Astronomy, University of Nevada, Las Vegas, 4505 S. Maryland Pkwy, Box 454002, Las Vegas, NV 89154-4002, USA}
\and Anders Johansen}
\institute{Lund Observatory, Department of Astronomy and Theoretical Physics, Lund University, Box 43, 22100 Lund, Sweden \\ \email{noemi.schaffer@astro.lu.se}}
\date{}

\abstract{The radial drift and diffusion of dust particles in protoplanetary disks affect both the opacity and temperature of such disks as well as the location and timing of planetesimal formation. In this paper, we present results of numerical simulations of particle-gas dynamics in protoplanetary disks that include dust grains with various size distributions. We consider three scenarios in terms of particle size ranges, one where the Stokes number $\tau_{\rm{s}} = 10^{-1} - 10^0$, one where $\tau_{\rm{s}} = 10^{-4} - 10^{-1}$ and finally one where $\tau_{\rm{s}} = 10^{-3} - 10^{0}$. Moreover, we consider both discrete and continuous distributions in particle size. In accordance with previous works we find in our multi-species simulations that different particle sizes interact via the gas and as a result their dynamics changes compared to the single-species case. The larger species trigger the streaming instability and create turbulence that drives the diffusion of the solid materials. We measure the radial equilibrium velocity of the system and find that the radial drift velocity of the large particles is reduced in the multi-species simulations and that the small particle species move on average outwards. We also vary the steepness of the size distribution, such that the exponent of the solid number density distribution, $\rm{d}\it{N}/\rm{d}\it{a} \propto \it{a^{-q}}$, is either $q = 3$ or $q = 4$. We overall find that the steepness of the size distribution and the discrete versus continuous approach have little impact on the results. The level of diffusion and drift rates are mainly dictated by the range of particle sizes. We measure the scale height of the particles and observe that small grains are stirred up well above the sedimented midplane layer where the large particles reside. Our measured diffusion and drift parameters can be used in coagulation models for planet formation as well as to understand relative mixing of the components of primitive meteorites (matrix, chondrules and CAIs) prior to inclusion in their parent bodies.}


\titlerunning{Streaming Instability of Multiple Particle Species}

\maketitle 

\section{Introduction}

 Protoplanets -- the building blocks of terrestrial planets as well as of the cores of gas giants, ice giants and super-Earths -- form by collisional growth from micrometer dust particles to bodies with sizes of thousands of kilometers. There are, however, several barriers that hinder the formation of planets. One barrier occurs early in the course of planet formation, since pebbles of millimeter-centimeter sizes do not stick efficiently when they collide. Their interaction typically results in bouncing, erosion and fragmentation \citep{Guttler10}.

Even if sticking would be perfect, radial drift limits particle growth. Due to the orbital velocity difference between the solid and gas components, the former feels a headwind and spirals in toward the central star. Meter sized solids have drift times of approximately a hundred years at 1 au \citep{Adachi1976, Weidenschilling1977, Youdin2010}. Particles grow maximally to a size where the growth time-scale equals the radial drift time-scale, resulting typically in centimeter sizes interior of 10 au and millimeter sizes in the outer disk \citep{Birnstiel2012, Lambrechts2014, Birnstiel2016}.

One way to overcome the radial drift barrier is to concentrate solids into clumps via the streaming instability \citep{Youdin2005, Youdin2007, Johansen2007}. Taking into account the back-reaction of the solids onto the gas, a small over-density of particles accelerates the surrounding gas to a higher speed. The radial drift of the given particle clump is thus reduced, since it feels less drag from the surrounding gas. At the same time, growth to  even larger clump size is possible by the accumulation of inwards drifting particles. Once the local solid concentration reaches the Roche density, planetesimals form through gravitational collapse \citep{Johansen2012}. The formation of filaments by the streaming instability occurs above a threshold metallicity slightly elevated relative to solar metallicity \citep{Johansen2009, Bai2010b, Carrera2015, Yang2017}. The metallicity may be increased either by gas photoevaporation \citep{Carrera2017} or by pile-up of drifting solids in the inner regions of the protoplanetary disk \citep{Drazkowska2016, Ida2016, Schoonenberg2017, Drazkowska2017, Gonzalez2017}.

In recent years, the amount of protoplanetary disk observations at different wavelengths has increased dramatically. Observations at millimeter and centimeter wavelengths reveal the presence of pebbles while scattered light observations probe the dust population.
Spatially resolved ALMA observations of young disks around Class 0 protostars show evidence of the presence of millimeter-sized pebbles, which hints to grain growth via coagulation already at an early stage of disk formation \citep{Gerin2017}. ALMA observations of the spectral index in the rings of the disk around HL Tau also suggests local grain growth up to centimeter sizes \citep{ALMAPartnership2015, Zhang2015}.
One of the closest observed systems with a protoplanetary disk, TW Hya, has been studied extensively by several different instruments as well. \cite{Menu2014} reviewed interferometric observations of the disk, from near-infrared up to centimeter wavelengths, that suggest the presence of both micron-sized dust and millimeter-centimeter-sized pebbles.
\cite{VanBoekel2016} presented scattered light observations of the TW Hya disk with the Spectro-Polari-meter High-contrast Exoplanet REsearch (SPHERE) instrument on the Very Large Telescope and \cite{Rapson2015} with the Gemini Planet Imager (GPI) on the Gemini South telescope; both these studies showed that the scattered light signal is dominated by dust. This indicates coexistence of solid particles of a range of sizes in young protoplanetary disks.

Even though grain growth is seen in observations, the majority of previous models involving the streaming instability focused on monodisperse particle populations. \cite{Bai2010a} filled this gap and considered a distribution of particle sizes to study the dynamics of materials in the midplane of protoplanetary disks. They assumed a discrete size distribution and saw that the radial drift velocity of all particle species is reduced compared to the one single species would have. Moreover, they showed that small grains tend to move outwards in the disk. \cite{Drazkowska2016} also implemented particle size dependent drift velocities into their numerical model and showed that this enhanches particle concentration. Using a fluid formalism, \cite{Laibe2014} provided analytical calculations considering multiple particle species and found the outward migration of small grains in protoplanetary disk settings as well. \cite{Johansen2007a} performed numerical simulations of planetesimal formation by the gravitational collapse of locally overdense regions, where they considered multiple solid sizes. They concluded that different particle sizes collapse into the same gravitationally bound system, despite the difference in their aerodynamic properties.

Our goal with this paper is to understand the interaction between the gas and the observed pebble population in protoplanetary disks.
We build on the work of \cite{Bai2010a} and study the dynamics of a wide range of particle sizes with the inclusion of mutual drag forces between the gas and solid species. 
We perform simulations with various particle size ranges and size distributions. More importantly, we study the evolution of both discrete and continuous systems in terms of particle size distributions and compare the results. In Sect. \ref{SectionNum} we discuss the numerical model and the dynamical equations. In Sect. \ref{SectionVelocity} we measure the equilibrium drift velocities in our models and discuss the effect of including a range of particle sizes on the radial drift velocities. In Sect. \ref{SectionScaleHeight} we address the evolution of the particle scale height throughout about 300 orbits and the influence of the particle size distribution on the degree of turbulence in the system. In Sect. \ref{DiffusionSection}, we analyze the radial and vertical diffusion of each solid species. In Sect. \ref{SectionImplication} we discuss potential implications of our results and finally in Sect. \ref{SectionDiscussion} we summarize our work.

\section{Numerical model} 
\label{SectionNum}

We model a local segment of protoplanetary disks in a shearing box, which co-rotates with Keplerian velocity, $\textbf{\textit{v}}_{\rm{K}}$, at an arbitrary distance, \textit{r}, from the central star. Our simulations are two-dimensional, in the radial-vertical plane, since this is sufficient for capturing the linear and non-linear evolution of the streaming instability \citep{Youdin2007, Johansen2007, Bai2010a}. For simplicity, we neglect magnetic effects. Moreover, we focus on the stage before planetesimal formation begins, and hence we neglect self-gravity of solid materials. 

\subsection{Dynamical equations}
In our models, the gas component is described as a fluid with density $\rho_{\rm{g}}$ and velocity $\textbf{\textit{u}}$ on a fixed grid. The gas velocity is measured relative to the Keplerian shear motion, which introduces additional advection terms in the dynamical equations \citep{Goldreich1965}. The equations that govern the motion of the gas are the continuity and momentum equations, respectively,

\begin{equation}
\frac{\partial \rho_{\rm{g}}}{\partial t} + \textbf{\textit{u}} \cdot \nabla \rho_{\rm{g}} - \frac{3}{2} \varOmega x \frac{\partial \rho_{\rm{g}}}{\partial y} = - \rho_{\rm{g}} \nabla \cdot \textbf{\textit{u}},
\label{EqGas1}
\end{equation}

\begin{equation}
\begin{split}
\frac{\partial \textbf{\textit{u}}}{\partial t} + (\textbf{\textit{u}} \cdot \nabla )  \textbf{\textit{ u}} & -  \frac{3}{2} \varOmega x \frac{\partial \textbf{\textit{u}}}{\partial y}  =   2 \varOmega u_y \hat{\textbf{\textit{x}}} - \frac{1}{2} \varOmega u_x \hat{\textbf{\textit{y}}} \\ & - \frac{\nabla P}{\rho_{\rm{g}}} \color{black}+ 2 \eta \varOmega^2 r \hat{\textbf{\textit{x}}} - \varOmega^2 z \hat{\textbf{\textit{z}}}- \frac{\epsilon}{\tau_{\rm{f}}}(\textbf{\textit{u}} - \textbf{\textit{v}}). 
\end{split}
\label{EqGas2}
\end{equation}

\noindent The third term on the left-hand side of Equation \eqref{EqGas1} and Equation \eqref{EqGas2} represents the background shear flow, where $\varOmega$ is the local Keplerian angular frequency. The first two terms on the right-hand side of Equation \eqref{EqGas2} account for the radial component of the stellar gravity and the centrifugal and Coriolis forces, while the third term results from the local pressure gradient. We assume an adiabatic equation of state with an adiabatic index of 5/3 and with an initial uniform adiabatic sound speed, $c_{\rm{s}}$, although in our simulations, $c_{\rm{s}}$ is close to being constant. The large-scale pressure gradient in the disk is represented by the fourth term, where 

\begin{equation}
\eta = - \frac{1}{2} \Bigg(\frac{H_{\rm{g}}}{r}\Bigg)^2 \frac{\partial \mbox{ln } P}{\partial \mbox{ln } r}
\label{Eq8.1}
\end{equation}

\noindent is the dimensionless parameter that sets the level of radial pressure support in the gas disk \citep{Nakagawa1986}. 
In Equation \eqref{Eq8.1} $H_{\rm{g}}$ is the gas scale height and $P$ is the pressure. The second-to-last term on the right-hand side of Equation \eqref{EqGas2} represents the vertical component of the stellar gravity, while the last term is responsible for the mutual drag forces between the solid and gas components. The velocity difference between the two components is described by $(\textbf{\textit{u}} - \textbf{\textit{v}})$, where $\textbf{\textit{v}}$ is the particle velocity. The friction time, $\tau_{\rm{f}}$, represents the time over which the gas drag changes the velocity of a given particle by a significant amount and $\epsilon = \rho_{\rm{p}} / \rho_{\rm{g}}$ is the ratio of solid and gas mass densities. 

The solid component is modeled as Lagrangian super-particles,  each of which represents a swarm of identical physical particles. The governing equations for each super-particle are 

\begin{equation}
\frac{\rm{d} \textbf{\textit{x}}_{\rm{p}}}{\rm{d}\it{t}} = -\frac{3}{2} \varOmega x_{\rm{p}} \hat{\textbf{\textit{y}}} + \textbf{\textit{v}}, 
\label{EqPar1}
\end{equation}

\begin{equation}
\frac{\rm{d} \textbf{\textit{v}}}{\rm{d}\it{t}} = \Bigg( 2 \varOmega v_y \hat{\textbf{\textit{x}}} - \frac{1}{2} \varOmega v_x \hat{\textbf{\textit{y}}} - \varOmega^2 z_{\rm{p}} \hat{\textbf{\textit{z}}} \Bigg) + \frac{\textbf{\textit{u}} - \textbf{\textit{v}}}{\tau_{\rm{f}}}.
\label{EqPar2}
\end{equation}

\noindent The position of the particle is $\textbf{\textit{x}}_{\rm{p}} = (x_{\rm{p}}, y_{\rm{p}}, z_{\rm{p}})$, while its velocity relative to the Keplerian shear is $\textbf{\textit{v}} = (v_{x}, v_{y}, v_{z})$ \citep{Youdin2007}. 

We consider solids that have constant friction time. Given that in the usual conditions of a protoplanetary disk particle sizes are smaller than the mean free path of gas molecules, we can translate the interaction between the gas and solids to be in the Epstein regime \citep{Weidenschilling1977}. In this case, friction time is

\begin{equation}
\tau_{\rm{f}} = \frac{\rho_{\bullet} a}{\rho_{\rm{g}} c_{\rm{s}}}.
\label{Eq1}
\end{equation}

\noindent Here, $\rho_{\bullet}$ and $a$ are the internal density and the radius of the solids. In the rest of the paper, we will use the dimensionless Stokes number, $\tau_{\rm{s}} = \tau_{\rm{f}} \varOmega$. In terms of the Stokes number and the gas surface density $\varSigma_{\rm{g}}$, Equation \eqref{Eq1} becomes

\begin{equation}
\tau_{\rm{s}} = \sqrt{2 \pi} \frac{\rho_{\bullet} a}{\varSigma_{\rm{g}}}.
\label{EqSt}
\end{equation}

\noindent We write the gas surface density as

\begin{equation}
\varSigma_{\rm{g}} (r) = 1000 \mbox{ g cm}^{-2} \Bigg( \frac{r}{\mbox{au}} \Bigg)^{-1} f_{\rm{g}},
\label{EqGasSurface}
\end{equation}

\noindent where $f_{\rm{g}}$ is a factor that measures gas depletion, with $f_{\rm{g}} = 1$ representing a relatively pristine disk with gas surface density comparable to the Minimum  Mass Solar Nebula of \cite{Hayashi1981}. Note, however, that we use a shallower power law index than the MMSN, motivated by observations of protoplanetary disks \citep{Andrews2009, Gorti2011, Bate2018}. Using Equation \eqref{EqSt} and Equation \eqref{EqGasSurface}, we can find the physical solid size that corresponds to a given Stokes number at different regions of a protoplanetary disk as

\begin{equation}
a \approx 400 \mbox{ cm }  \Bigg( \frac{r}{\mbox{au}} \Bigg) ^{-1}  \Bigg(\frac{\rho_{\bullet}}{1 \mbox{ g cm}^{-3}} \Bigg)^{-1} f_{\rm{g}} \tau_{\rm{s}}.
\label{EqSize}
\end{equation}

\noindent Assuming that $f_{\rm{g}} = 1$, Equation \eqref{EqSize} shows that with $\rho_{\bullet} = 1 \mbox{ g cm}^{-3}$ at 2.5 au $\tau_{\rm{s}} = 10^{-4}$ corresponds to solids of approximately 160 microns, while $\tau_{\rm{s}} = 1$ represents solids of about 160 centimeters.

\subsection{Initial conditions and boundary conditions}

Both the gas and solid components are initialized following a stratified density profile centered at the disk midplane. The particle scale height of all species is initially set to $H_{\rm{p}} = 0.015  H_{\rm{g}}$ as in \cite{Bai2010a}. 

In terms of initial gas and particle velocity, we have two setups. In the case of discretized particle size distribution, we initialize both the solid and gas drift speed according to the multi-species Nakagawa-Sekiya-Hayashi equilibrium solution \citep{Nakagawa1986, Tanaka2005, Bai2010a}. The velocities of $N$ particle species, which are described by $\gamma_x = (v_{1x}, v_{2x}, ..., v_{Nx})^{\rm{T}}$ and  $\gamma_y = (v_{1y}, v_{2y}, ..., v_{Ny})^{\rm{T}}$, satisfy the matrix equation

\begin{equation}
\left( \begin{array}{c} \gamma_x \\ \gamma_y \end{array} \right) = \eta v_{\rm{K}}  \begin{pmatrix} A & 2B \\ -B/2 & A \end{pmatrix} \left( \begin{array}{c} 0 \\ 1 \end{array} \right),
\label{Eq8}
\end{equation}

\noindent where $\eta v_{\rm{K}}$ represents the global radial pressure support felt by the gas (Equation \eqref{Eq8.1}). The (sub-)matrices \textit{A} and \textit{B} are defined as $A = \varLambda^{-1} (1 + \varGamma)B$ and $B = \{[\varLambda^{-1}(1+\varGamma)]^2 + 1 \}^{-1} \varLambda^{-1}$. Both are functions of the Stokes number $\tau_{i}$ and the dust-to-gas ratio $\epsilon_{i}$ for each species $i$, through $\varLambda = \mbox{diag}\{\tau_1, \tau_2, ..., \tau_{N} \}$ and $\varGamma = (\bm{\epsilon},\bm{\epsilon}, ..., \bm{\epsilon})^{\rm{T}}$, where $\bm{\epsilon} = (\epsilon_1, \epsilon_2, ..., \epsilon_{N})^{\rm{T}}$. It is useful to non-dimensionalise the pressure support in terms of $\varPi = \eta v_{\rm{K}} / c_{\rm{s}}$ \citep{Bai2010c}. \cite{Bai2010a} find that $\varPi \approx 0.05$ describes typical pressure support in the inner regions of protoplanetary disks, which is the value we apply in our models \citep[see also][]{Bitsch2015}. The initial gas velocity is then calculated by momentum conservation as 

\begin{equation}
\textbf{\textit{u}} = - \sum_j \epsilon_j \textbf{\textit{v}}_j - \eta v_{\rm{K}} \textbf{\textit{$\hat{\textbf{y}}$}}.
\label{Eq9}
\end{equation}

This method works well for discrete particle species, but becomes impractical for the case of a continuous particle size distribution. In that case, the gas component is therefore initialized with a sub-Keplerian velocity, such that $\textbf{\textit{u}}(t = 0) =  - \eta \textit{v}_{\rm{K}}\hat{\textit{\textbf{y}}}$ and the particle velocities are set to zero. 

The boundary conditions of the simulation domain  are reflecting in the vertical direction and periodic in radial direction.

\subsection{Numerical method}

To simulate the particle-gas systems, we use the Pencil Code\footnote{Publicly available at http://pencil-code.nordita.org/ }. It is a high-order finite difference code used for (magneto--)hydrodynamic flows in astrophysics and it uses third-order Runge-Kutta time integration \citep{Brandenburg2002}. Equations \eqref{EqGas1}, \eqref{EqGas2}, \eqref{EqPar1} and \eqref{EqPar2} are solved using explicit finite differencing. The coupling between the particle and gas components is achieved numerically using the Triangular Shaped Cloud (TSC) method, which is a second order particle-mesh scheme \citep{Youdin2007}.

\subsection{Simulation parameters}

\begin{table}
\centering
\caption{Simulation Parameters}
\label{Table1}
\begin{tabular}{l>{$}c<{$}>{$}c<{$}>{$}c<{$}>{$}c<{$}>{$}c<{$}}
\hline
\hline
Name & \tau_{\rm{s, min}} & \tau_{\rm{s, max}} & q & L_x \times L_z & N_x \times N_z \\  
 & & & & (H_{\rm{g}}) & \\
\hline
SI41-3-2-c/d & 10^{-4} &  10^{-1} & 3  & 0.2 \times 0.2 & 128 \times 128 \\
SI41-3-4-c & 10^{-4} &  10^{-1} & 3  & 0.4 \times 0.4 & 256 \times 256\\
SI30-3-2-c/d & 10^{-3} & 10^{ 0} & 3  & 0.2 \times 0.2 & 128 \times 128\\
SI30-3-4-c & 10^{-3} & 10^{ 0} & 3  & 0.4 \times 0.4 & 256 \times 256\\
SI30-3-8-c & 10^{-3} & 10^{ 0} & 3  & 0.8 \times 0.8 & 512 \times 512\\
SI10-3-2-c & 10^{-1} & 10^{ 0} & 3  & 0.2 \times 0.2 & 128 \times 128\\
\hline
SI41-4-2-c/d & 10^{-4} &  10^{-1} & 4  & 0.2 \times 0.2 & 128 \times 128 \\
SI41-4-2b-d & 10^{-4} &  10^{-1} & 4  & 0.2 \times 0.2 & 256 \times 256 \\
SI41-4-4-c/d & 10^{-4} &  10^{-1} & 4  & 0.4 \times 0.4 & 256 \times 256\\
SI30-4-2-c/d & 10^{-3} & 10^{ 0} & 4  & 0.2 \times 0.2 & 128 \times 128\\
SI30-4-4-c/d & 10^{-3} & 10^{ 0} & 4  & 0.4 \times 0.4 & 256 \times 256\\
SI10-4-2-d & 10^{-1} & 10^{ 0} & 4  & 0.2 \times 0.2 & 128 \times 128\\
\hline
SI41-4-2A-d \tablefootmark{a} & 10^{-4}   &  10^{-1} & 4  & 0.2 \times 0.2 & 128 \times 128\\
SI10-4-2B-d & 10^{-1} & 10^{ 0} & 4  & 0.2 \times 0.2 & 256 \times 256 \\
SI10-4-2C-d & 10^{-1} & 10^{ 0} & 4  & 0.2 \times 0.2  & 512 \times 512 \\
\hline
\end{tabular}
\tablefoottext{a}{Total particle number increased by factor of four}
\end{table}

Table \ref{Table1} lists the parameters that we used in our model. In the first column, we list the names of our runs. The nomenclature SI$MN$-$A$-$B$-$p$ describes some of the parameters of the given run. ''SI'' stands for the \textit{streaming instability}, while ''$M$'' and ''$N$'' describe the minimum and maximum particle sizes such that $\tau_{\rm{s, min}} = 10^{-M}$ and $\tau_{\rm{s, max}} = 10^{-N}$. Next, ''$A$'' describes the exponent, $q$, of the solid number density distribution, $\rm{d}\it{N} / \rm{d}\it{a} \propto \it{a^{-q}}$. We refer to a shallow distribution, when $q = 3$, and a steep distribution when $q = 4$. The next parameter ''$B$'' describes the size of the simulation box, such that $L_x = L_z = 0.B H_{\rm{g}}$. Lastly, ''$p$'' describes, whether the given run is continuous (c) or discrete (d) in terms of particle size distribution.

The second and third columns list the three different size ranges we used in our simulations. These Stokes number values correspond to solids ranging from about one hundred micrometers to about one meter in size (see Equation \eqref{EqSize}).

The fourth column in Table \ref{Table1} lists the slope of the size distribution of solids in our runs. The grain number density follows $\rm{d}\it{N} / \rm{d}\it{a} \propto \it{a^{-q}}$, where $N$ is the solid number density and $a$ is the solid size. Depending on physical processes considered, $q$ can take up different values. In the case of self-similar fragmentation cascade in a debris disk, $q = 3.5$ \citep{Dohnanyi1969, Williams1994}. This value also agrees with the analytic estimate of size distribution of small asteroids and grains in the interstellar medium \citep{Mathis1977}. In protoplanetary disks gas also plays an important role in the dynamics of solids. As the drag from the gas slows down small particles, coagulation can also take place, which changes the steady state particle distribution. \cite{Birnstiel2011} provided recipes for grain size distribution in the enveloping cases of different collision regimes. In this paper, we study the case of $q = 3$ (shallow distribution) and $q = 4$ (steep distribution) in order to stay consistent with various possibilities.

The fifth column in Table \ref{Table1} encapsulates the different box sizes that we used in our simulations. Our standard dimensions are 0.2 $\times$ 0.2 $H_{\rm{g}}$, which is sufficient for most of the runs. In some cases, we increased this size in order to study how the box size influences the results.

Finally, in the last column we list the number of grid points $N_{x}$ and $N_{z}$ in the $x$ and $z$ directions. As default, we used $N_x \times N_z \times N_{\rm{species}}$ particles, where $N_{\rm{species}}$ represents the number of particle sizes. Based on \cite{Bai2010b}, this is sufficient to capture the particle-gas dynamics with multi-sized particles. In addition, we used 128 $\times$ 128 grid cells ($N_x \times N_z$) in the radial and vertical directions. We performed convergence tests in terms of grid and particle resolutions. In run SI41-4-2A-d, we increased the number of particles by a factor of four from the default value. In runs SI10-4-2B-d and SI10-4-2C-d the number of cells was increased to 256 $\times$ 256 and 512 $\times$ 512, respectively.

The solid-to-gas ratio, $Z$, is another important factor in the outcome of the dynamics. This parameter is defined as $Z = \varSigma_{\rm{p}} / \varSigma_{\rm{g}}$, where 

\begin{equation}
\varSigma_{\rm{p,g}}  = \int_{- \infty}^{\infty} \rho_{\rm{p,g}}(r,z) dz. 
\label{Eq2}
\end{equation}

\noindent Here, $\varSigma_{\rm{p}}$ and $\varSigma_{\rm{g}}$ are the solid and gas column densities, respectively.  In our simulations, we used a value of $ Z = 0.01$, which does not result in efficient particle clumping \citep{Bai2010a, Carrera2015, Yang2017}. However, forming planetesimals is not the goal of this paper. Excluding strong solid concentration and thus planetesimal formation allows us to isolate the effect of the multiple particle sizes on the dynamics of the system.

\section{Equilibrium drift velocities and radial diffusion}
\label{SectionVelocity}

The relative radial motion of the solids and the gas is driven by their mutual drag force interactions. On one hand, the particles experience radial drift due to the headwind they feel from the sub-Keplerian gas. On the other hand, the relative motion between particles and gas also triggers the streaming instability and generates disk turbulence. As a consequence of the self-generated turbulence, in addition to radial drift, the solid material also experiences radial and vertical diffusion. In this section, we measure the radial velocity of particles once the system reaches the equilibrium state, after the vertical settling and diffusion are in balance.

\subsection{Discrete particle size distribution}
\label{SectionDiscreteVelocity}

Following \cite{Bai2010a} we began with a discrete particle size distribution. We first discretized particle sizes into bins of half a dex in Stokes number, $\tau_{\rm{s}}$, where each bin contained particles of the same $\tau_{\rm{s}}$. The velocities of both the particles and the gas were initialized according to the Nakagawa-Sekiya-Hayashi (NSH) solution \citep{Nakagawa1986, Bai2010a} given in Equation \eqref{Eq8} and Equation \eqref{Eq9}, respectively. To find the initial velocity profile of the two components at any given height, we needed to find an appropriate value for $\rm{\bm{\epsilon}}$ (and thus $\varGamma$). We calculated the local dust-to-gas ratio of each size bin as a function of height using the relationship

\begin{equation}
\epsilon_j (z) =  \frac{\rho_{j,0}\exp(-z^2 / 2H_{\rm{p}}^2)}{\rho_{\rm{g},0}\exp(-z^2 / 2H_{\rm{g}}^2 )},
\label{Eq10}
\end{equation}

\noindent where the gas density at the midplane is $\rho_{\rm{g},0}$, $z$ is the vertical coordinate, and $H_{\rm{p}}$ and $H_{\rm{g}}$ are the particle and gas scale heights, respectively. Here,

\begin{equation}
\rho_{j,0} =  \rho_{\rm{g},0} f_j Z  \frac{H_{\rm{g}}}{H_{\rm{p}}}
\label{Eq11}
\end{equation}
is the total mass density of particles of species $j$  in the midplane and the mass fraction $f_j$ is
\begin{equation}
f_j = \frac{\tau_{\rm{s},\it{j}}^{4-q}}{\sum_k \tau_{\rm{s},\it{k}}^{4-q}}.
\label{EqMassFrac}
\end{equation}

\noindent The Stokes number of different particle species is denoted by $ \tau_{\rm{s},\it{j}}$.

\subsubsection{Steep discrete size distribution ($q = 4$)}
\label{DiscSteep}
 
First, we considered equal particle mass in each logarithmic size bin. The size distribution parameter, \textit{q}, is related to the mass per logarithmic radius bin through

\begin{equation}
\frac{\mbox{d}M}{\mbox{dln }a} \propto a^{4-q},
\label{Eq14}
\end{equation}

\noindent where $M$ is the total solid mass. This implies that $q = 4$ corresponds to uniform solid mass per logarithmic size bin.

\begin{figure}[!t]
\centering
\includegraphics[width=1\columnwidth]{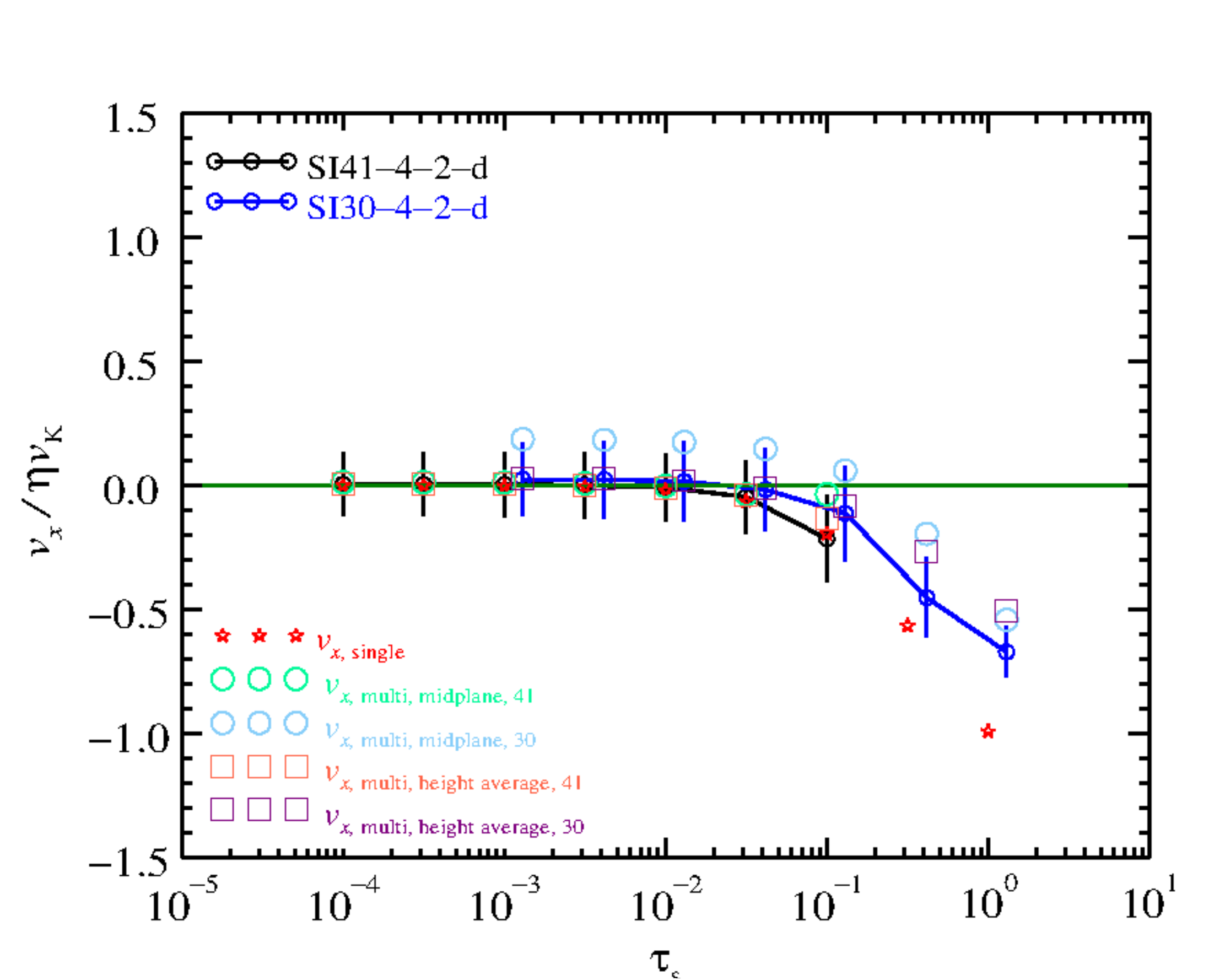}\caption{Mean equilibrium radial drift velocity as a function of Stokes number ($\tau_{\rm{s}}$) for runs SI41-4-2-d and SI30-4-2-d measured from the time of saturation. The vertical bars show the 1$\sigma$ variation of the radial velocity and are measures of the radial diffusion each species experiences. The red stars represent the single species equilibrium solutions. The green and blue circles show the multi-species radial equilibrium velocity near the midplane. The purple and red squares show the height averaged multi-species solution under the assumption that the particle density of each species is distributed uniformly within its scale height. Run SI30-4-2-d was shifted horizontally for illustration purposes.\label{disc:f1}}
\end{figure}

Figure \ref{disc:f1} shows the mean radial equilibrium velocity of runs SI41-4-2-d and SI30-4-2-d. The measurement was started once the particles settled to a quasi-steady scale height, at which point the vertical settling and diffusion are in balance. All particle velocities in the rest of the paper are normalized by $\eta v_{\rm{K}}$, which is the difference between the gas and Keplerian velocities in the absence of particles. 

\begin{figure}[t!]
\centering
\includegraphics[width=1\columnwidth]{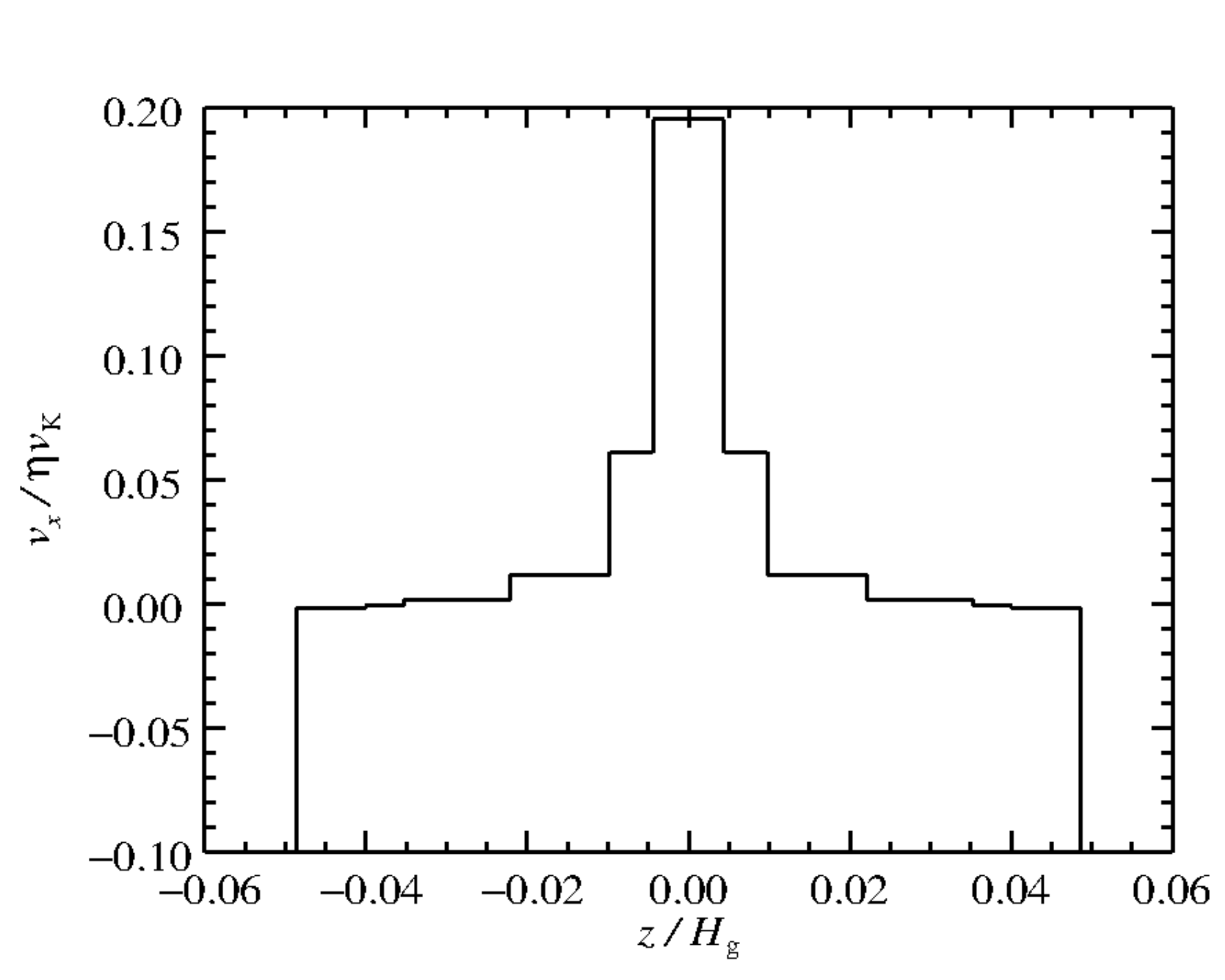}
\caption{Theoretical equilibrium radial drift velocity of species $\tau_{\rm{s}}=10^{-3}$ in run SI30-4-2-d with respect to height. The drift velocity is highest in the two bins around the midplane and is close to the gas velocity in the absence of large particles further away from the midplane.}
\label{vx_sp1}
\end{figure}

The vertical lines show the $1 \sigma$ variation. They represent the level of fluctuations around the mean drift velocity and correspond to the radial diffusion the given particle species experiences. The diffusion of all sizes is relatively uniform in the case of run SI41-4-2-d. In the case of run SI30-4-2-d the level of diffusion is lower for the largest particles $\tau_{\rm{s}} = 10^{-0.5}$ and $\tau_{\rm{s}} = 10^{0}$ than for the smaller sizes. This may be due to the fact that particles of these sizes are the least coupled to the gas and are not affected by its turbulence to a high degree. We discuss the effect of turbulence in more detail in Sect. \ref{DiffusionSection}.

In general, Fig. \ref{disc:f1} shows that the radial drift velocity decreases (becomes more negative) with increasing Stokes number. Most interestingly, the smallest particles have positive mean drift velocities, meaning that they move outwards. We compared the measured velocities in Fig. \ref{disc:f1} with the theoretical single species solution (red stars) using the NSH solution given by \cite{Nakagawa1986}

\begin{equation}
v_x = - \frac{2 \tau_{\rm{s}}}{(1 + \epsilon)^2 + \tau_{\rm{s}}^2} \eta v_{\rm{K}}.
\label{Eq15}
\end{equation}

\noindent We see that the radial drift velocities in our multi-species runs are reduced, especially for the larger Stokes numbers. The deviation of our results compared to the single species NSH solution is due to the fact that different particle species interact with each other through the gas. As large particles drift in, the gas moves out due to momentum conservation. Because small particles are more tightly coupled to the gas, they are entrained in its motion, and contribute to the acceleration of the gas to closer to the Keplerian speed. Thus, the larger particles drift in at reduced velocities compared to the single species case, since they feel a weaker headwind from the gas. At the same time, the particles with $\tau_{\rm{s}}  \gtrsim 10^{-2}$ trigger the streaming instability most readily and create turbulence in the surrounding gas, which diffuses the solid materials \citep{Bai2010a}. 

We compared our measurements with the theoretical multi-species solution (see Equation \eqref{Eq8}) using two methods. The first method gives the radial equilibrium velocity near the midplane. Here, we used the midplane dust-to-gas ratio of each species calculated under the assumption that the vertical particle distribution is approximately Gaussian. The midplane density for the particles is

\begin{equation}
\rho_{\rm{ p}} (z=0) = \frac{\varSigma_{\rm{p}}}{\sqrt{2 \pi} H_{\rm{p}}},
\label{Eq16}
\end{equation} 

\noindent where $\varSigma_{\rm{p}}$ is the column mass density of the particles.
Thus, the local dust-to-gas ratio for the $j$th particle bin at the midplane is

\begin{equation}
\epsilon_j = \frac{\varSigma_{\rm{p,}\it{j}}}{\sqrt{2 \pi} \ \overline{H}_{\rm{p,}\it{j}}} \frac{1}{\rho_{\rm{g,0}}},
\label{EqDusttoGasSimp}
\end{equation}

\noindent where $\overline{H}_{\rm{p,}\it{j}}$ is the average particle scale height in the $j$th bin and $\rho_{\rm{g,0}}$ is the gas density at the midplane. We can calculate the midplane dust-to-gas ratio in a given size bin using the global dust-to-gas ratio and the mass fraction per bin (see Equation \ref{EqMassFrac}), such that

\begin{equation}
\epsilon_j = \frac{Z f_j H_{\rm{g}}}{\overline{H}_{\rm{p,}\it{j}}}.
\label{Eq17}
\end{equation}

As seen in Fig. \ref{disc:f1}, the theoretical radial equilibrium velocity near the midplane agrees well in the case of run SI41-4-2-d (green circles), but is noticeably more positive than our measured drift velocities in the case of run SI30-4-2-d (blue circles). This is likely the result of the fact that we used the midplane dust-to-gas ratio, which is an overestimate for the solid loading further away from the midplane.

In the second method of calculating the analytic multi-species solution, we used the height-averaged radial equilibrium velocity. In this case, we calculated the drift velocity of each species as a function of height assuming particles of each species are uniformly distributed within its own scale height. Then the dust-to-gas ratio for species $j$ is given by

\begin{align}
& \epsilon_j(|z| < \overline{H}_{\rm{p,}\it{j}}) = \frac{\sqrt{2\pi} Z f_j H_{\rm{g}}}{2 \overline{H}_{\rm{p,}\it{j}}}, \label{EqPiecewise} \\ 
& \epsilon_j(|z| > \overline{H}_{\rm{p,}\it{j}}) = 0, \nonumber
\end{align}

\noindent where $\overline{H}_{\rm{p,}\it{j}}$ is the time-averaged scale height of the particle species. Figure \ref{vx_sp1} shows the equlibrium velocity of the $\tau_{\rm{s}}=10^{-3}$ species as a function of height in run SI30-4-2-d. The drift velocity is largest in two bins around the midplane and is close to the gas velocity in the absence of particles at larger heights. We then averaged the radial drift velocity over height by weighing with the dust-to-gas ratio in each bin \citep{Bai2010a}. As seen in Fig. \ref{disc:f1} (red and purples squares), the height averaged radial equilibrium velocity provides a significantly better match to the results of our simulations.

\begin{figure}[!t]
\centering
\includegraphics[width=1\columnwidth]{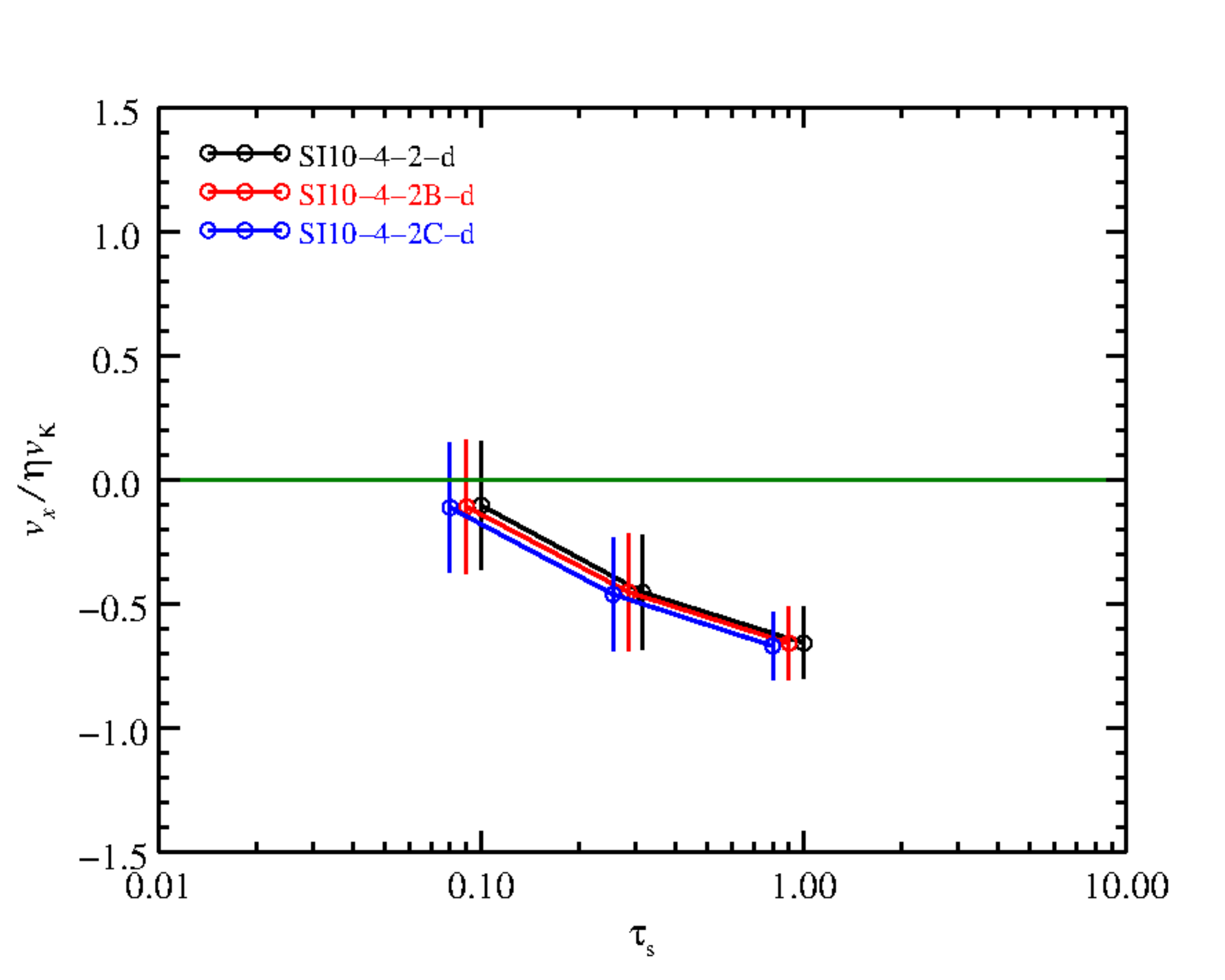}\caption{Convergence of equilibrium radial velocity of different particle sizes against grid resolution with runs SI10-4-2-d, SI10-4-2B-d  and SI10-4-2C-d. Both the mean drift velocities (circles) and the level of diffusion (vertical bars) agree between the cases of $N_x \times N_z = 128 \times 128$, $256 \times 256$ and $512 \times 512$. This implies that our results are not dependent on numerical resolution. Runs SI10-4-2B-d and SI10-4-2C-d were shifted horizontally for better visibility.\label{disc:f2}}
\end{figure}

\begin{figure}[!h]
\centering
\includegraphics[width=1\columnwidth]{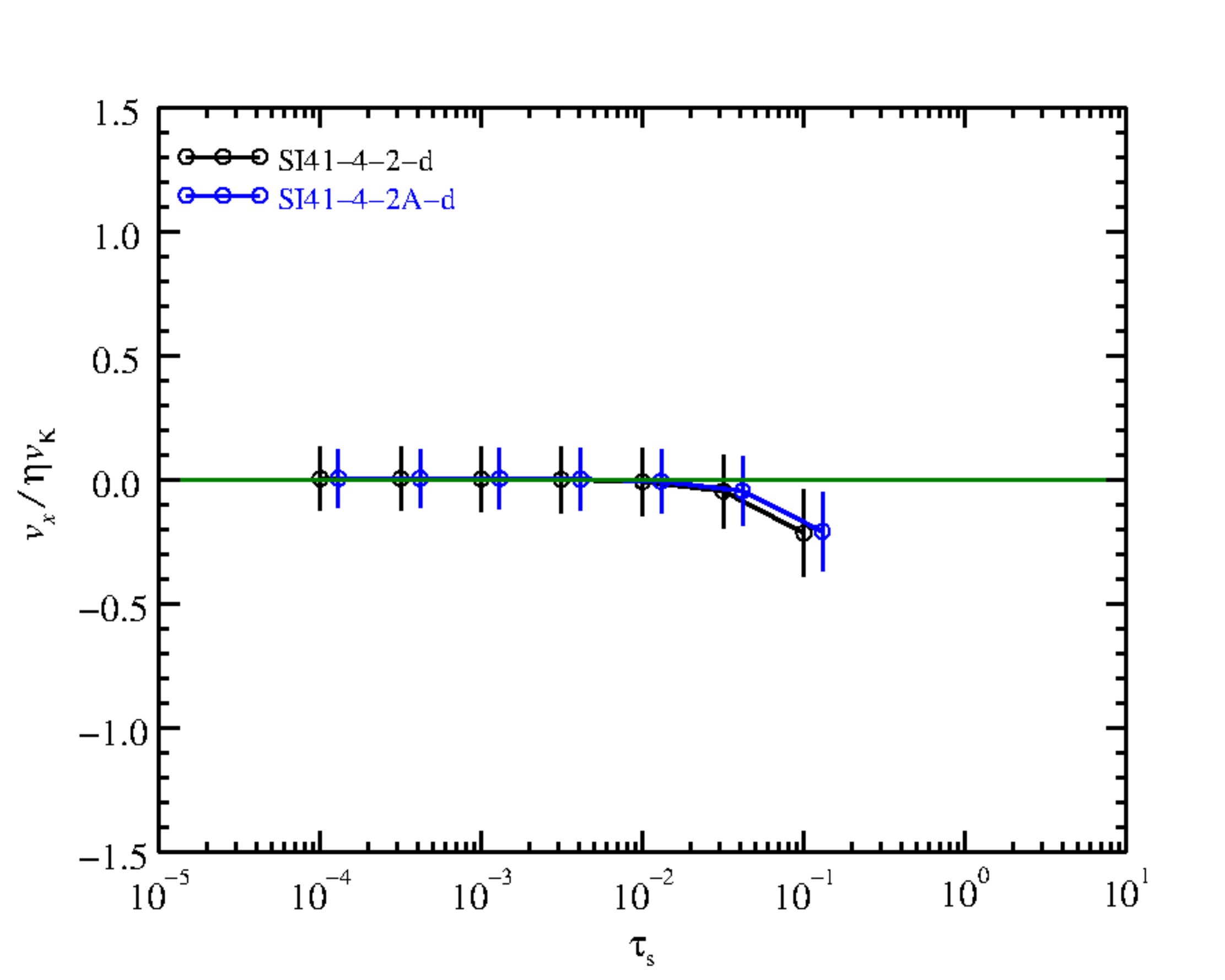}\caption{Convergence of equilibrium radial velocity of different particle sizes against particle number with runs SI41-4-2-d and SI41-4-2A-d. The mean drift velocities and the 1$\sigma$ variation agree well, indicating that the number of particles does not influence our results. In the case of run SI41-4-2A-d, the symbols were shifted horizontally for comparison purposes.\label{dics:f3}}
\end{figure}

We conducted convergence tests to ensure that numerical resolution does not influence our results. Figure \ref{disc:f2} shows the radial equilibrium velocity of particles in runs SI10-4-2-d, SI10-4-2B-d and SI10-4-2C-d. In all three cases, the simulation box size was $L_{x} = L_{z} = 0.2 H_{\rm{g}}$, but the number of grid cells was successively increased from $128 \times 128$ to $256 \times 256$ and finally to $512 \times 512$. The equilibrium velocity agrees in all three resolutions. Thus, it appears the grid resolution does not have a significant effect on our results. 

We also increased the total particle number by a factor of four, to check for convergence. Figure \ref{dics:f3} shows that indeed, the equilibrium radial velocity is not appreciably affected by an increase in particle number.

\subsubsection{Shallow discrete size distribution ($q = 3$)}

We also considered a size distribution with $q=3$, with which $\rm{d}\it{M}/\rm{d} \mbox{ln } \it{a} \propto \it{a}$. Compared to the previous case, we now had more mass in particles with larger Stokes numbers, thus there were more particles available to play an active role in the streaming instability. As before, to initialize the system we used Equation \eqref{Eq8} and Equation \eqref{Eq9}. 

\begin{figure}
\centering
\includegraphics[width=1\columnwidth]{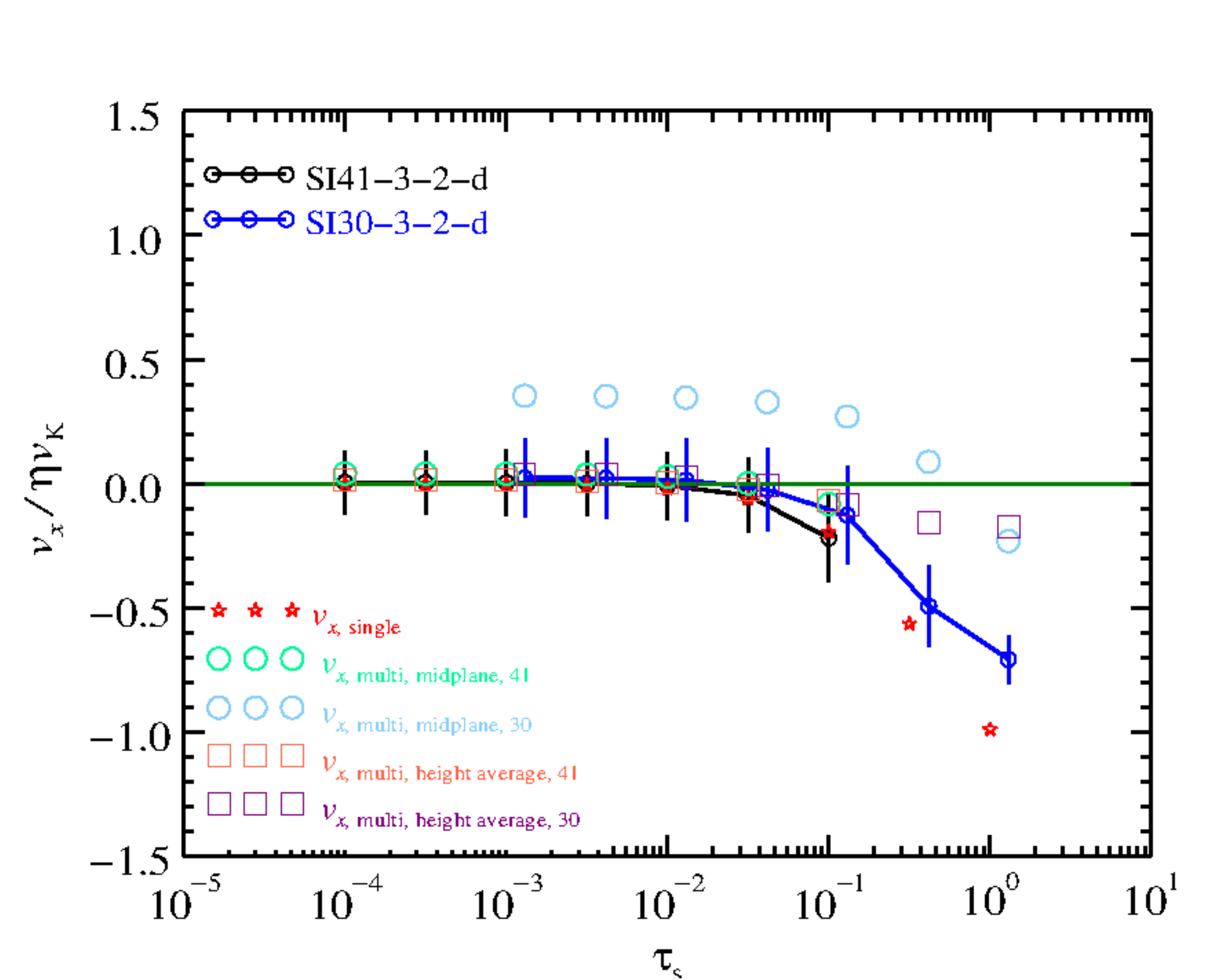}\caption{Mean equilibrium radial drift velocity as a function of $\tau_{\rm{s}}$ for runs SI41-3-2-d and SI30-3-2-d measured from the time of saturation. The color scheme agrees with that of Fig. \ref{disc:f1}. Run SI30-3-2-d was shifted horizontally for illustration purposes. 
\label{disc:f4}}
\end{figure}

Figure \ref{disc:f4} shows the radial equilibrium velocity of runs SI30-3-2-d and SI41-3-2-d. Similar to the previous case, where all size bins had the same solid mass (runs SI30-4-2-d and SI41-4-2-d) presented in Fig. \ref{disc:f1}, all particle species have unique drift velocities. The small particles have positive drift velocities and that of the larger species is reduced relative to the single species solution (marked as red stars). The $1 \sigma$ bars around the mean drift velocity represent the level of diffusion each species experiences. The radial diffusion is smallest for particles with $\tau_{\rm{s}} > 10^{-0.5}$, since these are the least coupled to the gas and therefore the least influenced by its turbulence.

In Fig. \ref{disc:f4} we again plot the theoretical multi-species solution using the two methods presented in the previous section. The green and blue circles correspond to the multi-species radial equilibrium velocity near the midplane. This method produces velocity values that agree well with our measurements in the case of run SI41-3-2-d, but deviate in the case of run SI30-3-2-d. The reason for the deviation is, again, that using the midplane dust-to-gas ratio (see Equation \ref{Eq17}) overestimates the solid loading further away from the midplane thus the drift velocity of the smaller species as well.

We compare the theoretical multi-species solution using the second method as well. The red and purple squares in Fig. \ref{disc:f4} correspond to the height averaged solution, where we assumed that the particle density of each species is distributed uniformly within its own scale height and used Equation \ref{EqPiecewise} to calculate the dust-to-gas ratio as a function of height. We then weighed the velocities by the particle density in each height bin. Figure \ref{disc:f4} shows that this method provides a significantly better match to our numerical measurements, especially in the case of the smaller species.

\subsection{Continuous particle size distribution}

So far we have considered discretized particle size distribution. We now turn to a more realistic case, where each particle in a given run has a unique size, i.e. particle sizes are distributed in a quasi-continuous way. In this case, particle velocities are initialized to be zero and the gas component is set to have an initial sub-Keplerian velocity of $u_y = - \varPi c_{\rm{s}}$, where $\varPi = 0.05$. 

\subsubsection{Steep continuous size distribution ($q = 4$)}

\begin{figure*}[!th]
  \begin{subfigure}[b]{0.5\textwidth}
\centering
\includegraphics[width=1\columnwidth]{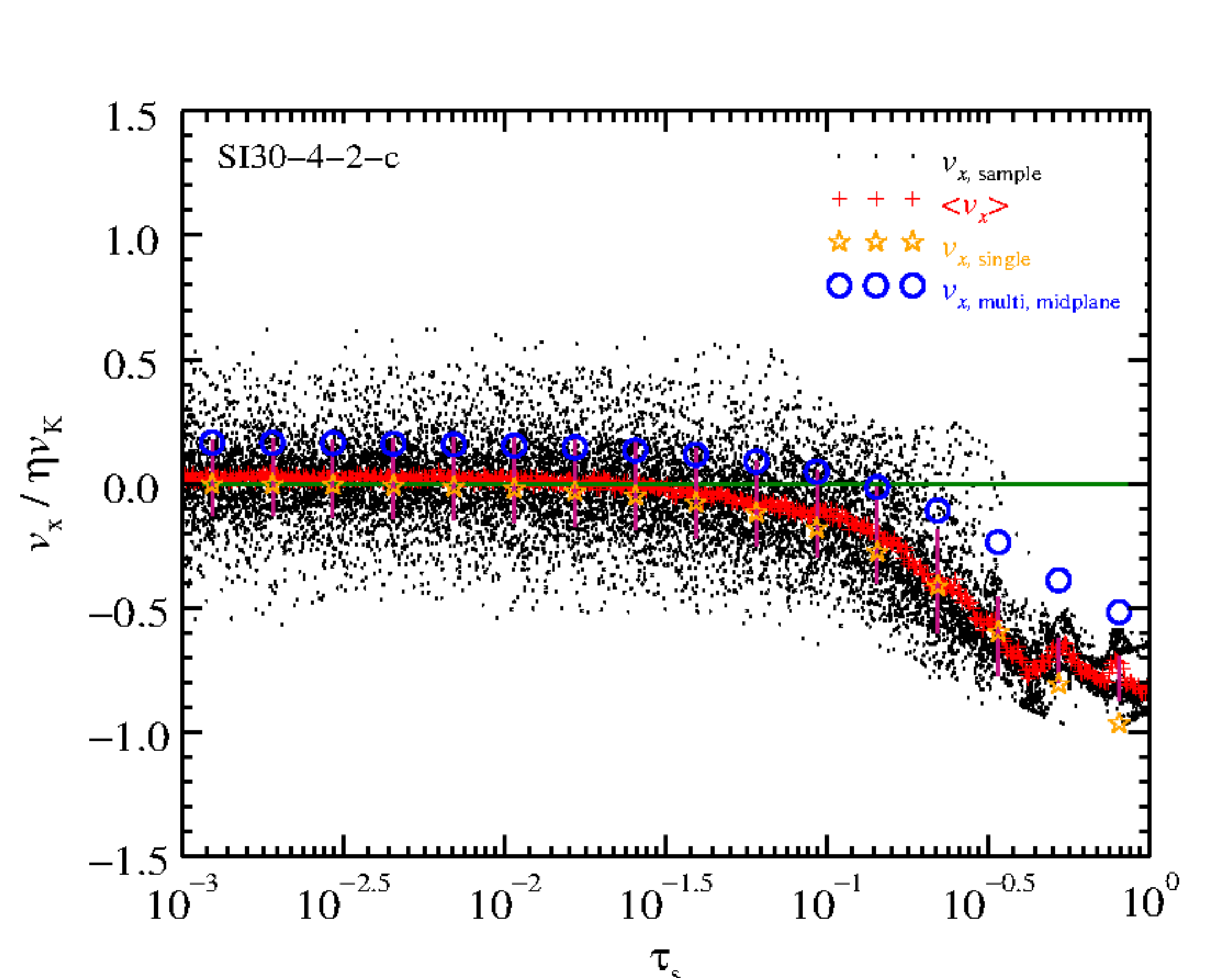}\caption{\label{fig:f1}}
\end{subfigure}
  \begin{subfigure}[b]{0.5\textwidth}
\centering
\includegraphics[width=1\columnwidth]{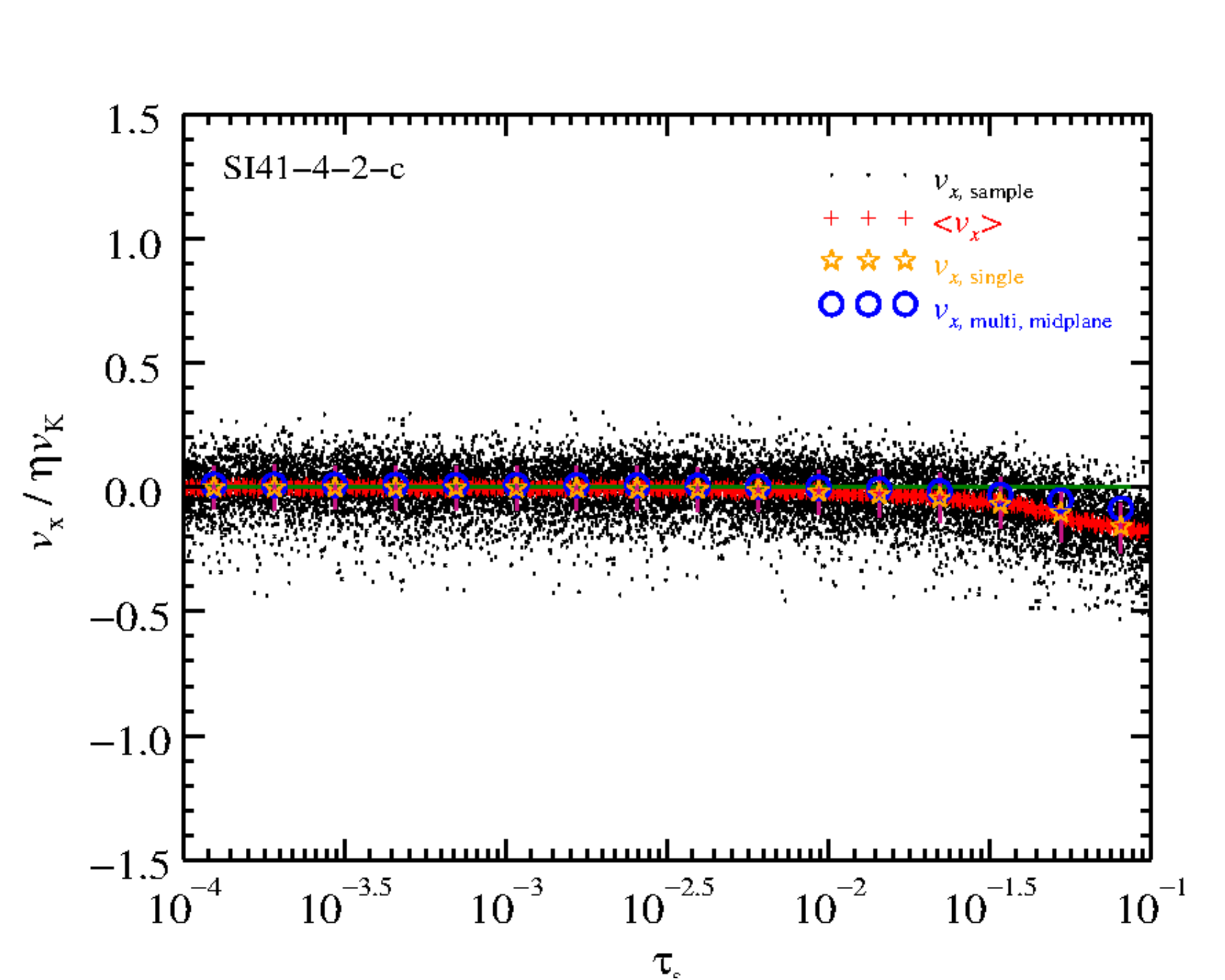}\caption{\label{fig:f2}}
\end{subfigure}\\
  \begin{subfigure}[b]{0.5\textwidth}
\centering
\includegraphics[width=1\columnwidth]
{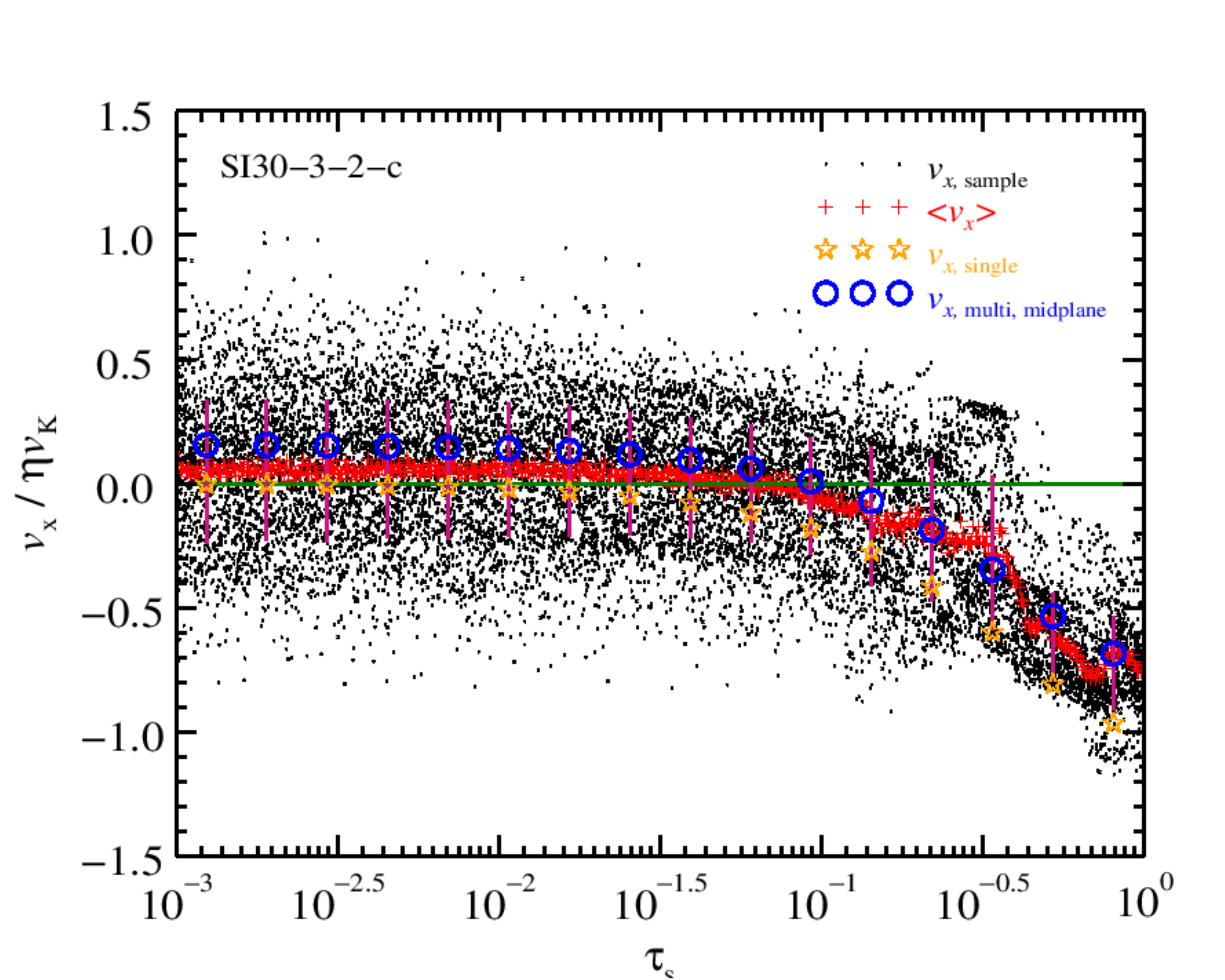}\caption{\label{fig:f3}}
\end{subfigure}
  \begin{subfigure}[b]{0.5\textwidth}
\centering
\includegraphics[width=1\columnwidth]{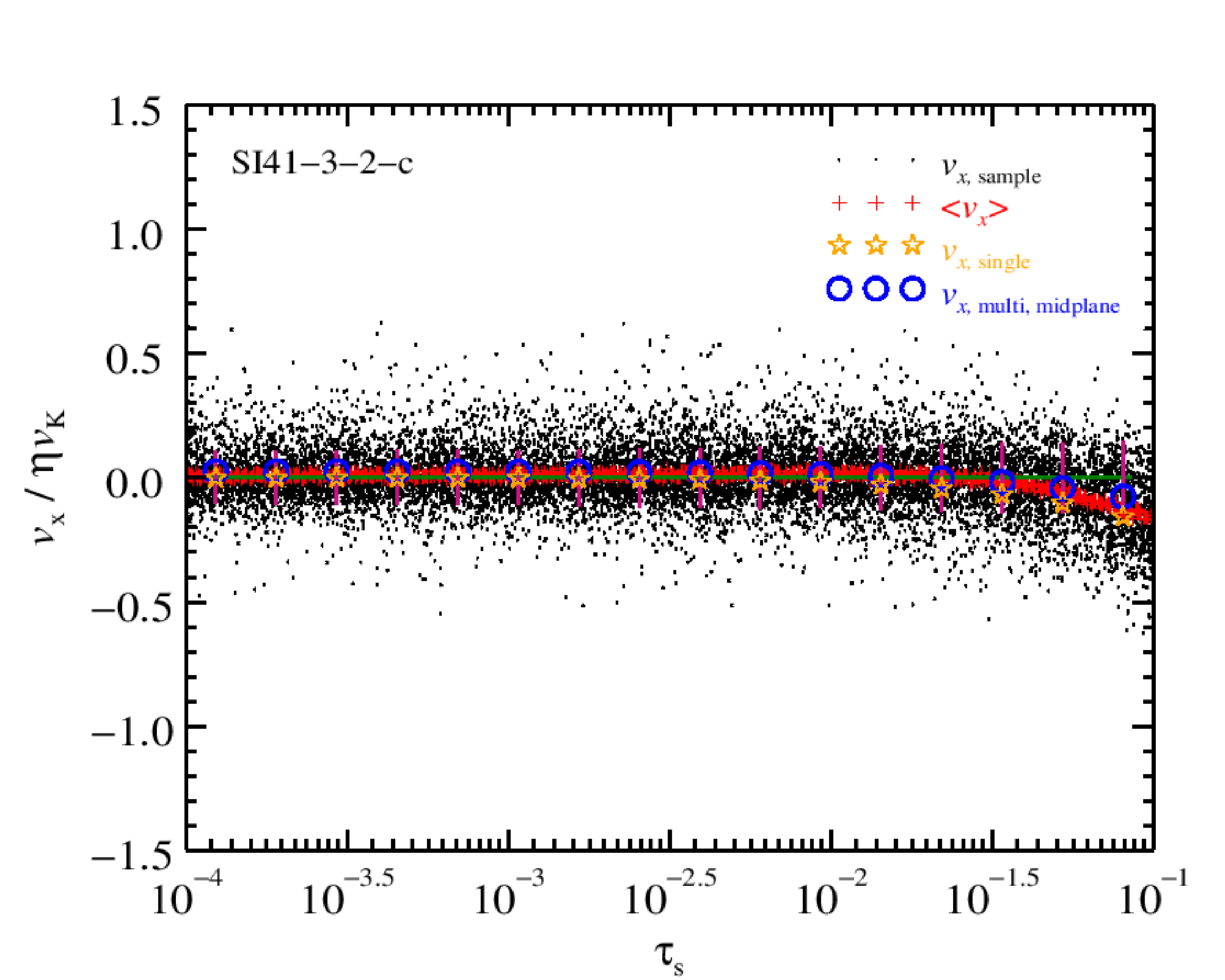}\caption{\label{fig:f4}}
\end{subfigure}
\caption{Radial drift velocity of runs SI30-4-2-c, SI41-4-2-c, SI30-3-2-c and SI41-3-2-c in terms of Stokes number ($\tau_{\rm{s}}$) at $t \approx 1000 \varOmega^{-1}$. Individual black points show the radial drift velocities of a random subset of approximately 14,000 particles. The particle size distribution is continuous, such that each particle has a unique size and drift velocity. Each red cross shows the mean drift velocity of two hundred particles with similar $\tau_{\rm{s}}$, which decreases with increasing particle size. The dispersion of points around this mean shows the level of diffusion the particles experience, which goes down with increasing particle size as well. The purple bars show the $1 \sigma$ variation around the mean velocity. The blue circles show the radial equilibrium velocity near the midplane for each particle species grouped into 16 bins using the multi-species NSH solution in Equation \eqref{Eq8}. The theoretical velocity values provide a relatively good fit to the mean drift velocity for small particle sizes but are somewhat off for large Stokes numbers.}
\label{VelocityCont304}
\end{figure*}

We re-visited the case of $q = 4$, such that $\rm{d}\it{M}/\rm{d} \mbox{ln } \it{a}$ is constant, and studied the same systems as in Sect. \ref{DiscSteep}, but using a quasi-continuous particle size distribution. Here each particle swarm was given a unique mass (per unit azimuthal length) according to the expression 

\begin{equation}
m_{\rm{p}, \mathit{j}} = \frac{M_{\rm{p}, \mathit{j}}}{N_{\rm{p},\mathit{j}}} = \frac{Z \rho_{0} \sqrt{2 \pi} H_{\rm{g}} L_{x} f_\mathit{j}}{N_{\rm{p},\mathit{j}}}, 
\end{equation}

\noindent where ${M_{\rm{p}, \mathit{j}}}$ is the total mass of a given particle species and $N_{\rm{p},\mathit{j}}$ is the number of particles of a given species. 

The resulting radial drift velocity of a random subset of all particles is shown in Fig. \ref{fig:f1}. In contrast to Fig. \ref{disc:f1}, now each particle has a unique velocity distributed around the mean shown with red. The level of diffusion is indicated by the scattering of the individual (black) drift velocity points around the mean. Similarly to the discrete case shown with blue in Fig. \ref{disc:f1}, the level of diffusion is smallest for the largest particles, since these are the least coupled to the gas \citep[see also][]{Youdin2007a}.

In Fig. \ref{fig:f1}, the blue circles represent the theoretical velocity values of all particles grouped into 16 bins calculated using the multi-species NSH solution (Equation \eqref{Eq8}). For this, we used the midplane dust-to-gas ratio in the given size bin calculated under the assumption that the vertical particle distribution is approximately Gaussian. The dust-to-gas ratio in the midplane of each size bin is calculated using Equation \ref{Eq17}. Since we used the midplane equilibrium radial velocity as comparison to our measurements, the theoretical velocities somewhat overestimate the numerical results, especially in the case of run SI30-4-2-c.

Figure \ref{fig:f2} shows the case of the smaller particle sizes, namely run SI41-4-2-c. Compared to the result in Fig. \ref{fig:f1} the particles are more concentrated around the mean velocity. This means that the level of diffusion is lower on average, indicating that the particles experience less turbulence. This is due to the fact that now, the largest particles have size $\tau_{\rm{s}} \leq 10^{-1}$ and more tightly coupled particles generate lower levels of turbulence. Thus the amount of particles that are large enough to drive the streaming instability is lower compared to the  case of run SI30-4-2-c \citep{Bai2010a}.

\subsubsection{Shallow continuous size distribution ($q = 3$)}
We also considered the case of the shallow particle size distribution. Figure \ref{fig:f3} shows a randomly selected sample of particles corresponding to run SI30-3-2-c. The mean velocity is positive for the smaller particles. Compared to the same run with $q = 4$ (Fig. \ref{fig:f1}) the diffusion of the smaller particles has increased.

We compared this result to the case with smaller Stokes numbers, namely run SI41-3-2-c. The drift velocities in Fig. \ref{fig:f4} are less dispersed around the mean, unlike in the case of the larger particles in Fig. \ref{fig:f3}. Also, both the theoretical and the mean velocities for larger sizes are smaller. As in the case of the steep size distribution, the theoretical midplane velocities overestimate somewhat our measurements, especially in the case of run SI30-3-2-c.

Comparing Fig. \ref{fig:f1} with Fig. \ref{fig:f3} and Fig.  \ref{fig:f2} with Fig. \ref{fig:f4}, it is clear that the use of either $q = 3$ or $q = 4$ does not have a large effect on the radial drift velocity. The level of turbulence appears to be higher if the particle size distribution is shallow, as shown by the larger dispersion in Fig. \ref{fig:f3} and Fig. \ref{fig:f4}.

\subsection{Comparison of particle size distributions}

In order to understand how parameters such as the exponent of the size distribution ($q$) and the manner in which the particles are distributed (discrete or continuous) influence the level of turbulence, we compared the 1$\sigma$ variation around the mean equilibrium radial drift velocities.

Whether the particles are distributed in a discrete or continuous way, is not expected to influence the results significantly. The only difference that stems from the two distribution methods is that in the latter case, each particle has a unique size between $\tau_{\rm{s, min}}$ and $\tau_{\rm{s, max}}$, wheres in the former case, the particle sizes are binned into discrete sizes.

\subsubsection{Comparison of steep and shallow particle size distributions}

\begin{figure*}[!th]
\centering 
\begin{minipage}[t]{1\columnwidth}
  \centering
  \includegraphics[width=\columnwidth]{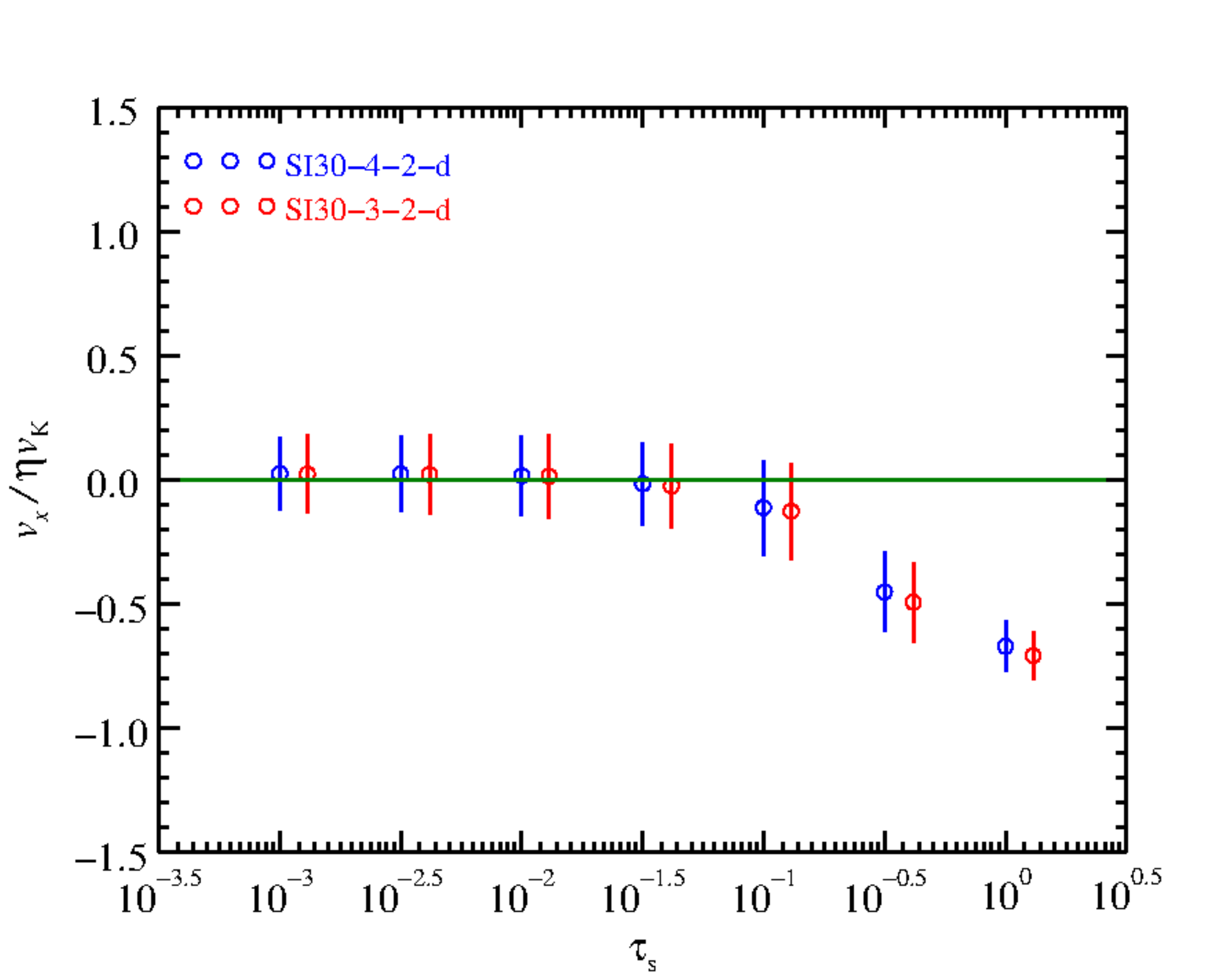}
  \caption{Mean equilibrium radial drift velocity and the 1$\sigma$ variation of the radial velocity as a function of Stokes number ($\tau_{\rm{s}}$) for runs SI30-4-2-d and SI30-3-2-d measured from the time when the saturated state is reached. The level of diffusion is similar in both runs.}\label{v_3_vs_4:f1}
\end{minipage}\hfill
\begin{minipage}[t]{1\columnwidth}
  \centering
  \includegraphics[width=\linewidth]{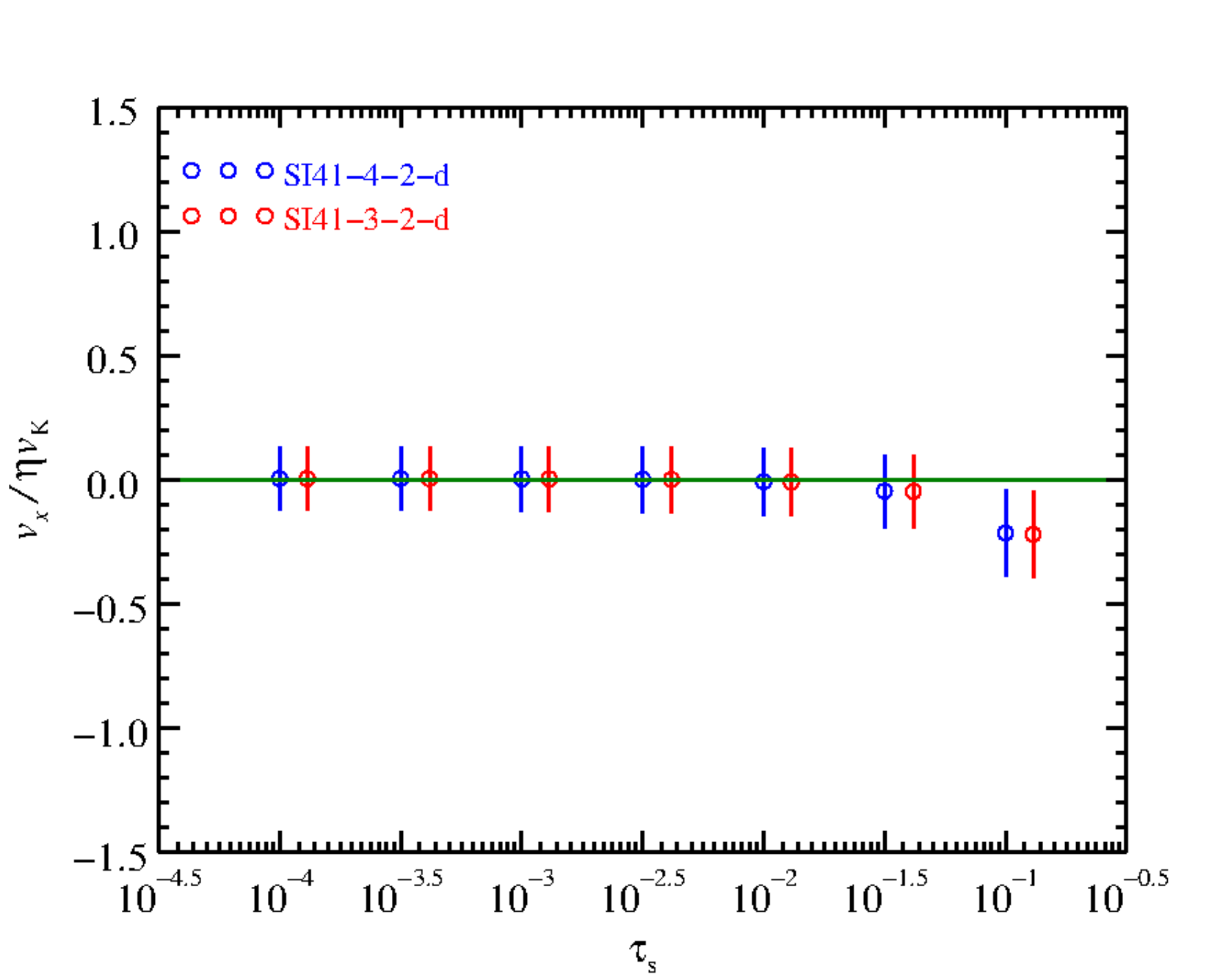}
    \caption{Mean equilibrium radial drift velocity and the 1$\sigma$ variation of the radial velocity as a function of Stokes number ($\tau_{\rm{s}}$) for runs SI41-4-2-d and SI41-3-2-d measured from the time when the saturated state is reached. The level of turbulence does not appear to be affected by the choice of the $q$ parameter.} \label{v_3_vs_4:f2}
\end{minipage}\hfill
\end{figure*}

In Fig. \ref{v_3_vs_4:f1}, we present the mean equilibrium drift velocity and the 1$\sigma$ limits in the case of run SI30-4-2-d and SI30-3-2-d. As the figure shows, there is no significant difference between the fluctuations around the mean velocities whether the particle size distribution is steep ($q=4$) or shallow ($q=3$). We see a similar tendency in Fig. \ref{v_3_vs_4:f2}, where we compare runs SI41-4-2-d and SI41-3-2-d and show that the 1$\sigma$ variation around the mean equilibrium drift velocities is similar for both particle size distributions.

\begin{figure*}[!th]
\centering 
\begin{minipage}[t]{1\columnwidth}
  \centering
  \includegraphics[width=\columnwidth]{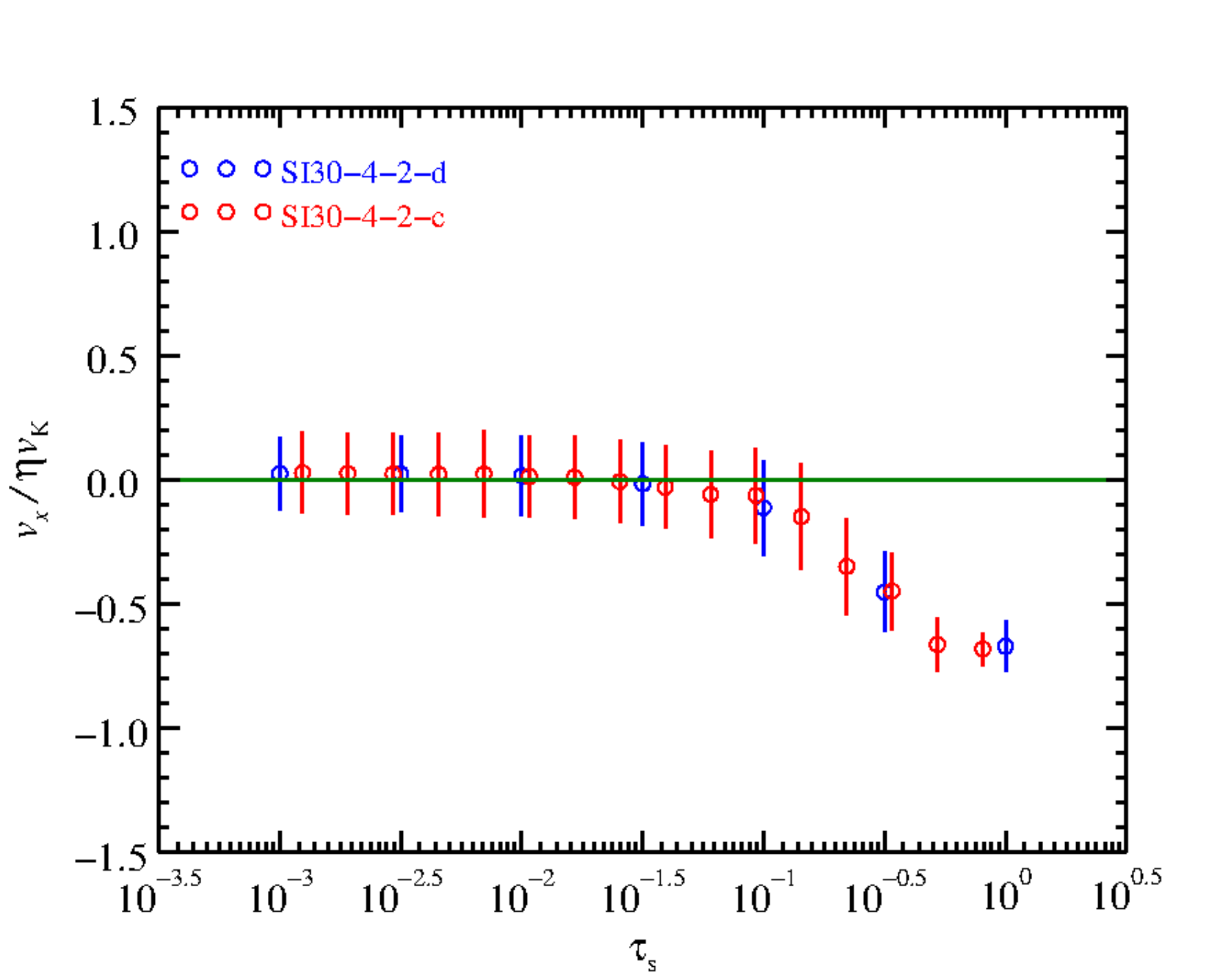}
  \caption{Mean equilibrium radial drift velocity and the 1$\sigma$ variation of the radial velocity as a function of Stokes number ($\tau_{\rm{s}}$) for runs SI30-4-2-d and SI30-4-2-c measured from the time when the saturated state is reached. In the case of run SI30-4-2-d, the velocities are averaged to the end of the simulation, while in the case of run SI30-4-2-c, they are averaged over 15 orbits.} \label{v_d_vs_c:f1}
\end{minipage}\hfill
\begin{minipage}[t]{1\columnwidth}
  \centering
  \includegraphics[width=\linewidth]{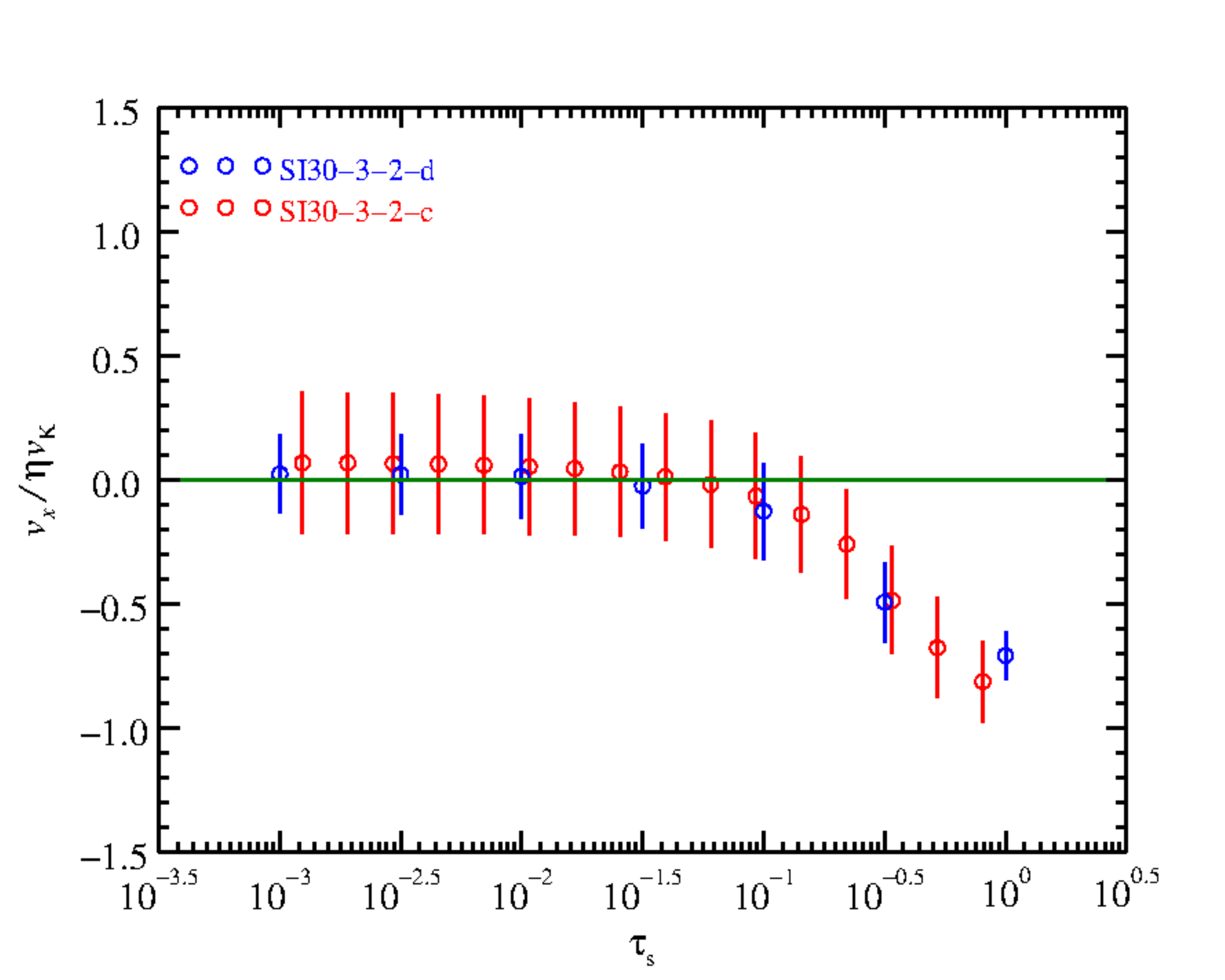}
    \caption{Mean equilibrium radial drift velocity and the 1$\sigma$ variation of the radial velocity as a function of Stokes number ($\tau_{\rm{s}}$) for runs SI30-3-2-d and SI30-3-2-c measured from the time when the saturated state is reached. In the case of run SI30-3-2-d, the velocities are averaged to the end of the simulation, while in the case of run SI30-3-2-c, they are averaged over 15 orbits.} \label{v_d_vs_c:f2}
\end{minipage}\hfill
\end{figure*}

\subsubsection{Comparison of discrete and continuous particle size distributions}

Figure \ref{v_d_vs_c:f1} shows the mean drift velocities and the 1$\sigma$ variation around them in the case of run SI30-4-2-d and SI30-4-2-c. The particle sizes are binned into 16 bins for the continuous run for illustration purposes and the velocities are calculated over 15 orbits. Since the particle sizes in the two runs do not agree completely, the comparison is not entirely straightforward. However, there is no significant difference between the two models in Fig. \ref{v_d_vs_c:f1}.

In Fig. \ref{v_d_vs_c:f2}, however there is some difference between the 1$\sigma$ limits of runs SI30-3-2-d and SI30-3-2-c. The run where the particles are distributed continuously shows about a factor of two stronger turbulence than the discrete case.

\section{Particle scale height}\label{SectionScaleHeight}

The mutual drag between the solid and gas components creates turbulence that stirs the particles \citep{Johansen2007}. Once the vertical settling and the  diffusion are in balance, the system reaches a saturated state, and the particles reach a quasi-steady scale height \citep{Johansen2009, Bai2010a, Yang2014}. By measuring the scale height to which each particle species saturates, we can indirectly assess the level of diffusion that they experience and thus the strength of turbulence driven by the streaming instability.

\begin{figure*}[!ht]
\centering
\begin{minipage}[t]{1\columnwidth}
  \centering
  \includegraphics[width=\columnwidth]{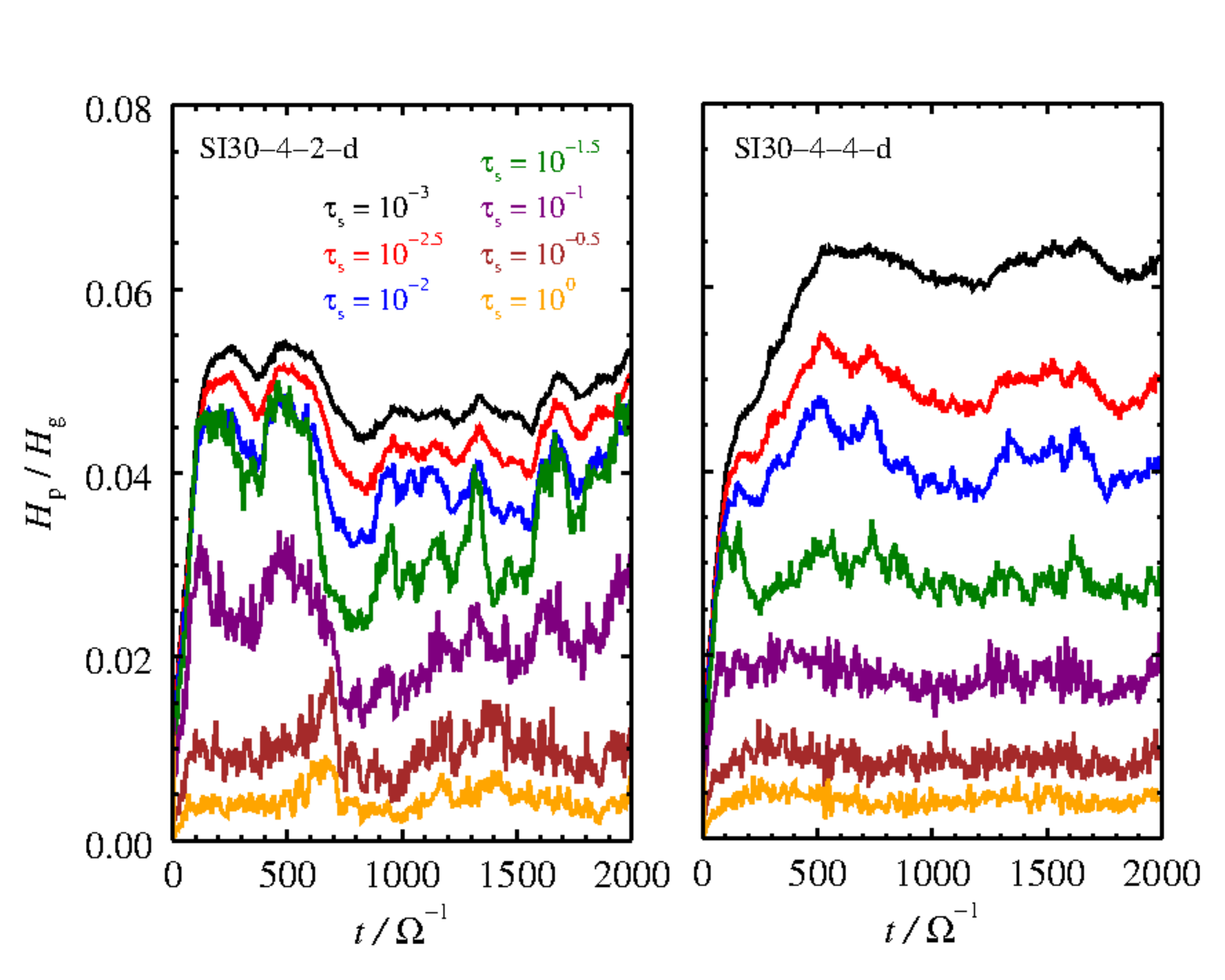}
  \caption{Particle scale height evolution of runs SI30-4-2-d and SI30-4-4-d for about 300 orbits. Different particle sizes (marked with different colors) saturate to different scale heights, in such a way that $H_{\rm{p}}$ increases with decreasing Stokes number. The plot on the left-hand side shows the evolution of the run with the larger Stokes numbers ($\tau_{\rm{s}} = 10^{-3} - 10^{0}$) in a $0.2 H_{\rm{g}} \times 0.2 H_{\rm{g}}$ simulation box. The plot on the right-hand side shows the convergence of the scale height for the larger species in a box that is twice the size of the original. The scale-height is relatively similar for the two box sizes, except for the two smallest particle size bins that exhibit an increased scale-height when the box size is enlarged.  \label{hp:1}}
\end{minipage}\hfill
\begin{minipage}[t]{1\columnwidth}
  \centering
  \includegraphics[width=\columnwidth]{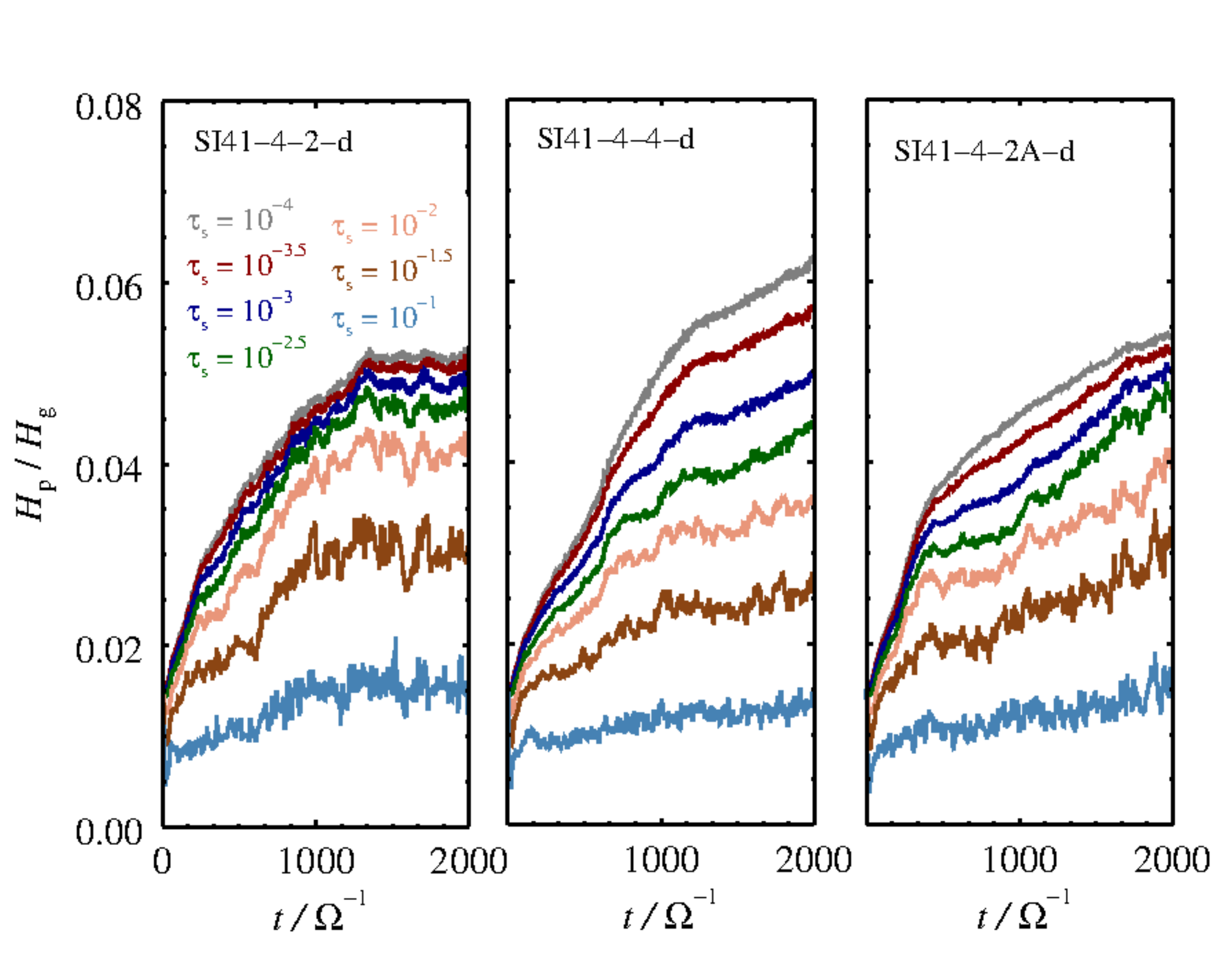}
  \caption{Particle scale height evolution of runs SI41-4-2-d, SI41-4-4-d, and SI41-4-2A-d, where the participating particles are $\tau_{\rm{s}} = 10^{-4} - 10^{-1}$. The plot on the left corresponds to the simulation in a box of $0.2 H_{\rm{g}} \times 0.2 H_{\rm{g}}$. The middle plot shows the evolution of the same system in a $0.4 H_{\rm{g}} \times 0.4 H_{\rm{g}}$ box, where the particle scale heights saturate to different values relative to the smaller box. In the right-most plot the total particle number was increased by a factor of four from the model in the left-most panel. In runs SI41-4-4-d and SI41-4-2A-d the particle subdisks do not saturate to a constant height within the simulation time. \label{hp:2}}
\end{minipage}\hfill
\end{figure*}

\begin{figure*}[!ht]
\centering 
\begin{minipage}[t]{1\columnwidth}
  \centering
  \includegraphics[width=\columnwidth]{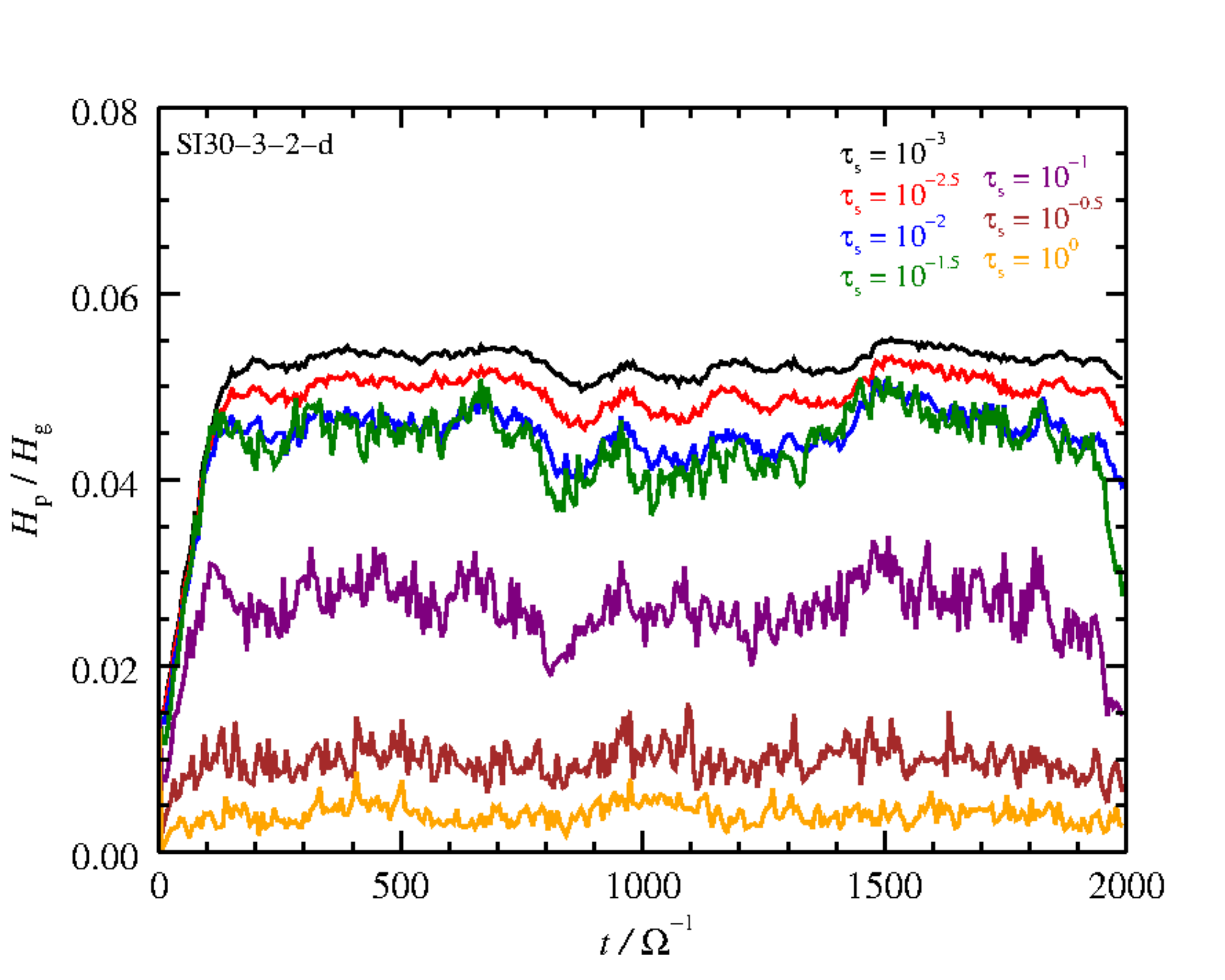}
  \caption{Particle scale height evolution of run SI30-3-2-d in a $0.2 H_{\rm{g}} \times 0.2 H_{\rm{g}}$ simulation box. Compared to the left-hand side of Fig. \ref{hp:1}, the  particle species saturate to similar scale heights but do not show fluctuations of the same magnitude with time. \label{hp:3}}
\end{minipage}\hfill
\begin{minipage}[t]{1\columnwidth}
  \centering
  \includegraphics[width=\linewidth]{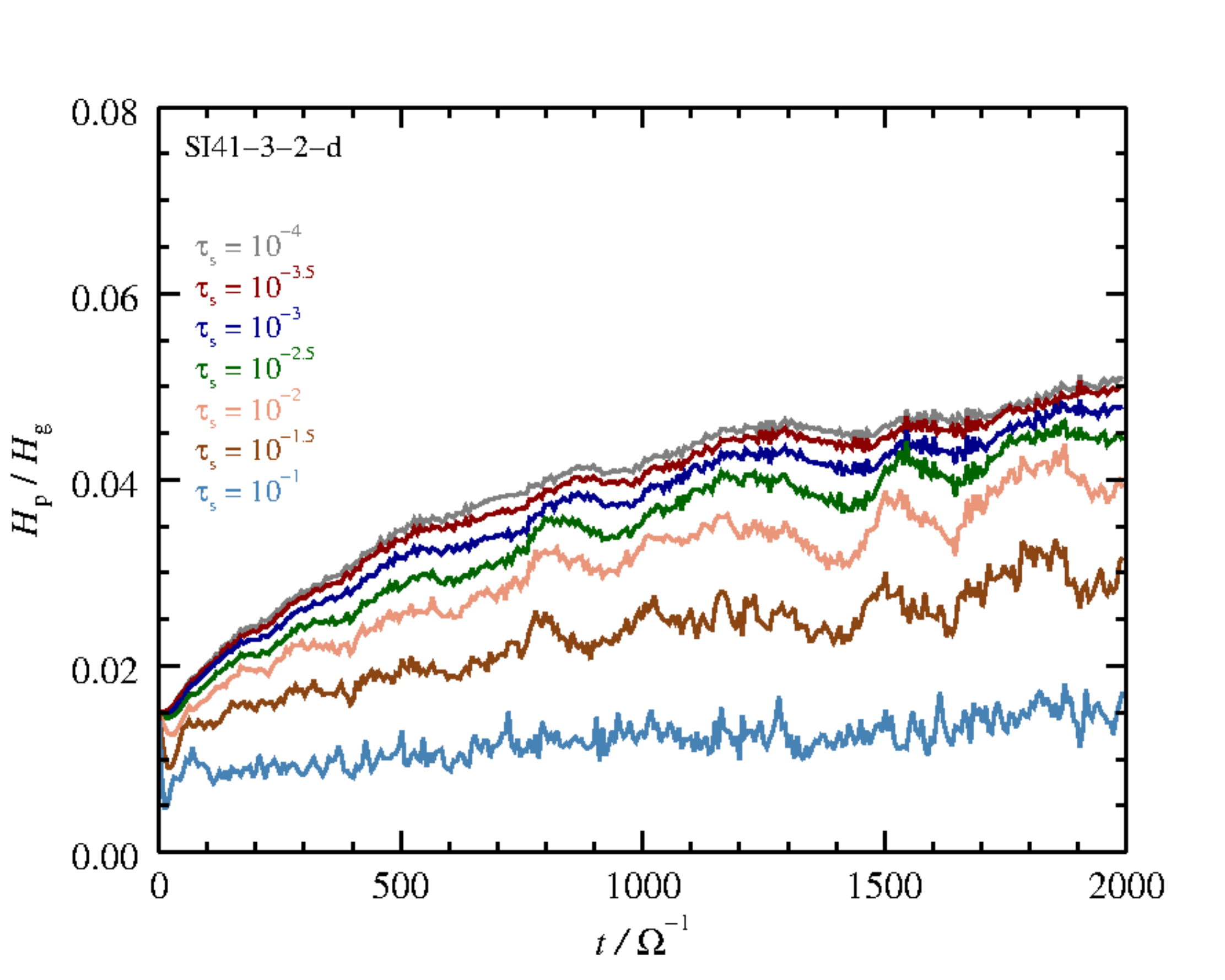}
    \caption{Particle scale height evolution of run SI41-3-2-d in a $0.2 H_{\rm{g}} \times 0.2 H_{\rm{g}}$ simulation box. The particle species do not reach a constant scale height within the simulation time. Compared to the plot on the left-hand side of Fig. \ref{hp:2}, where the power-law index of the distribution $q = 4$, the particle layers extend to similar scale heights. \label{hp:4}}
\end{minipage}\hfill
\end{figure*}

We calculate the scale height of each species as the square root of the particle height variance, $\sigma_z^2$, thus

\begin{equation}
H_{\rm{p}} = \sqrt{\sigma_z^2} = \sqrt{\langle z^2_{\rm{p}} \rangle - \langle z_{\rm{p}} \rangle ^2}.
\label{Eq18}
\end{equation}

\subsection{Discrete size distribution}

\begin{figure*}[!ht]
\centering
\begin{minipage}[t]{1\columnwidth}
  \centering
  \includegraphics[width=\columnwidth]{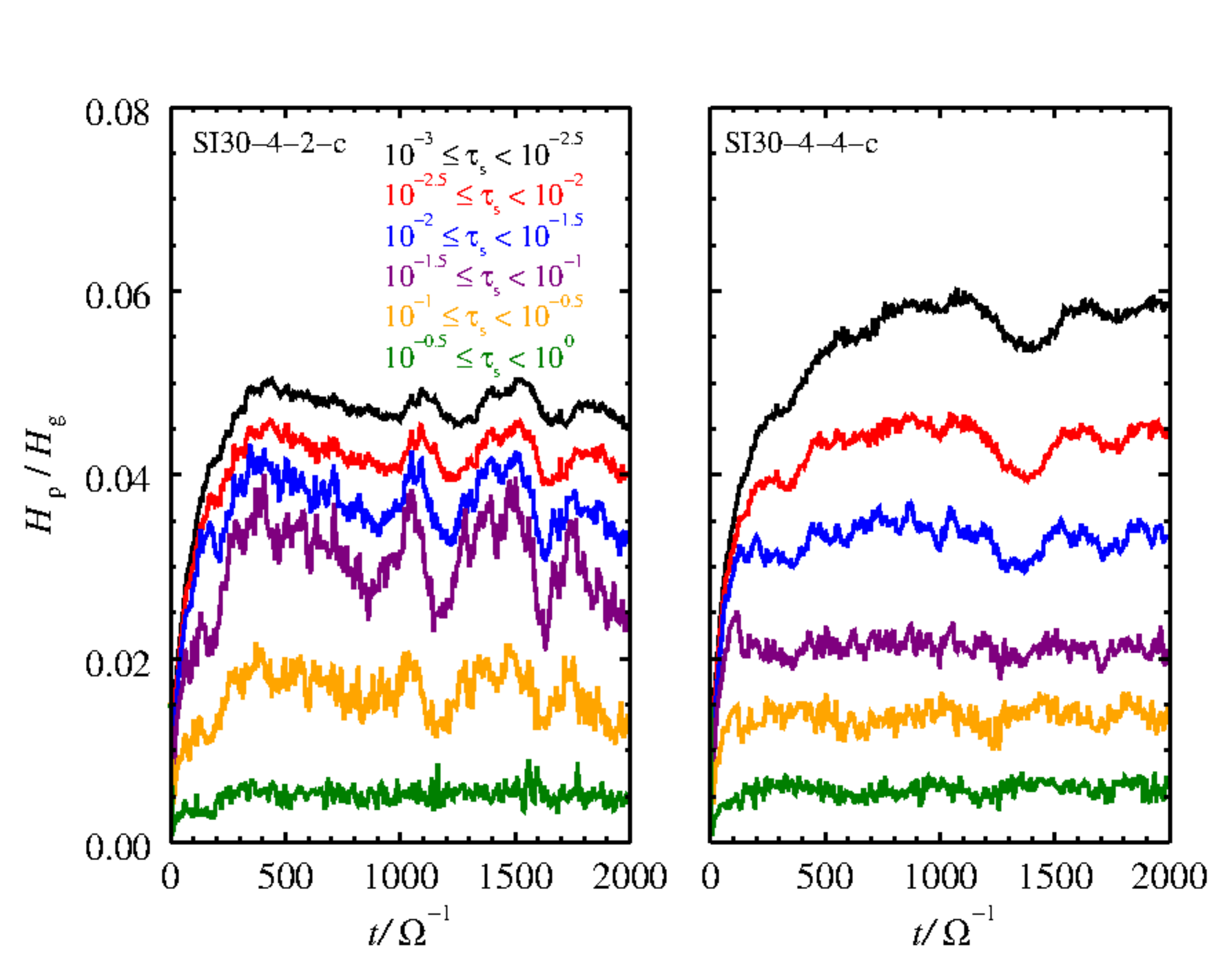}
  \caption{Particle scale height evolution of run SI30-4-2-c and run SI30-4-4-c. Each particle in the simulation has a unique size within the given size range. We grouped the particles into six size bins for illustration purposes. To ensure that the diffusion of the small particles is not limited by the simulation box, we performed the same simulation in a $0.4 H_{\rm{g}} \times 0.4 H_{\rm{g}}$ box in the right panel. The scale heights of the large particles converge while the smallest species saturate to a larger $H_{\rm{p}}$ than in the left panel. \label{hp:5}}
\end{minipage}\hfill
\begin{minipage}[t]{1\columnwidth}
  \centering
  \includegraphics[width=\columnwidth]{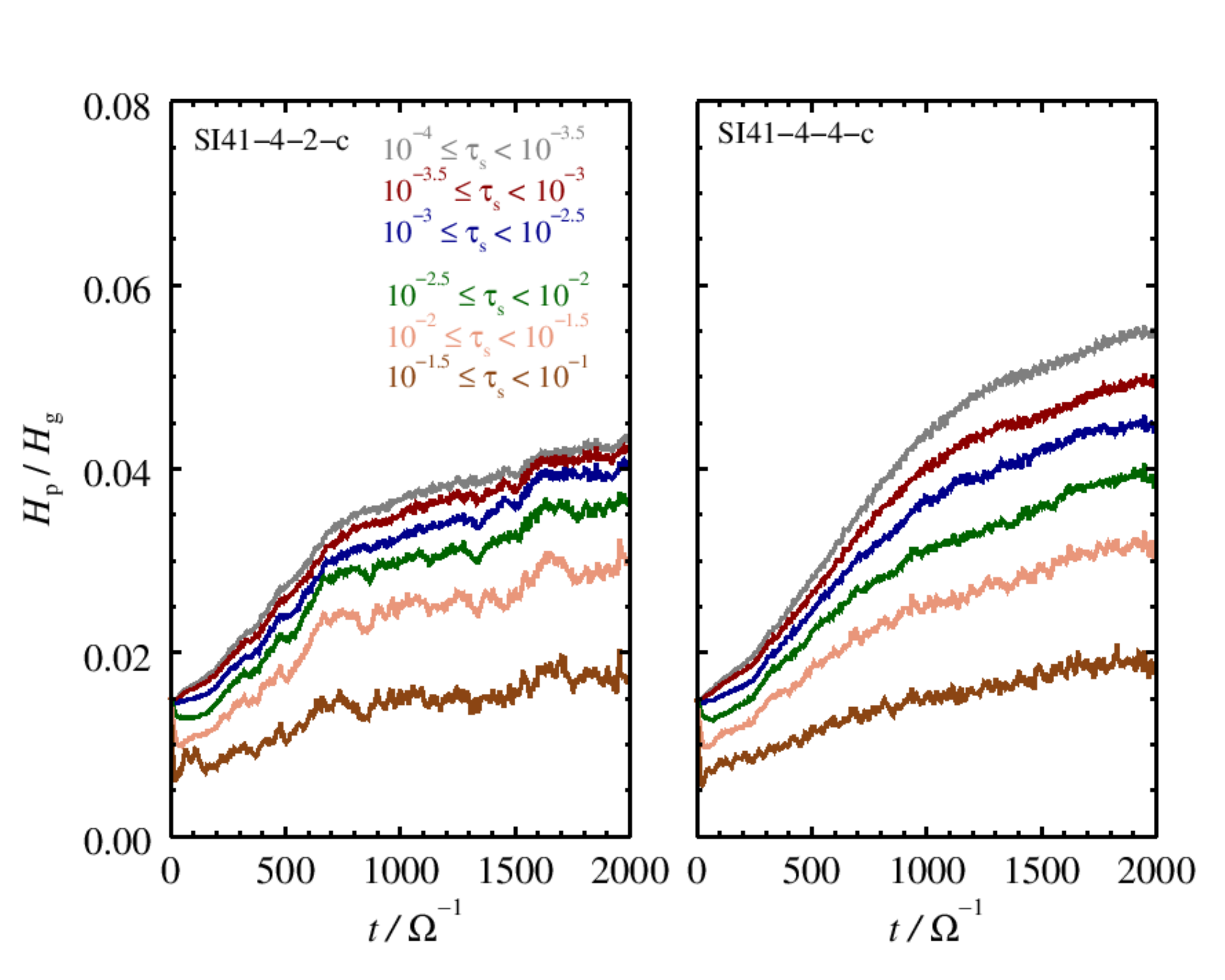}
  \caption{Particle scale height evolution of run SI41-4-2-c and run SI41-4-4-c. The scale height of all species does not saturate within the simulation time in either the $0.2 H_{\rm{g}} \times 0.2 H_{\rm{g}}$ (left), nor the $0.4 H_{\rm{g}} \times 0.4 H_{\rm{g}}$ box (right). \label{hp:6}}
\end{minipage}\hfill
\end{figure*}

\begin{figure*}[!t]
\centering
\begin{minipage}[t]{1\columnwidth}
  \centering
  \includegraphics[width=\columnwidth]{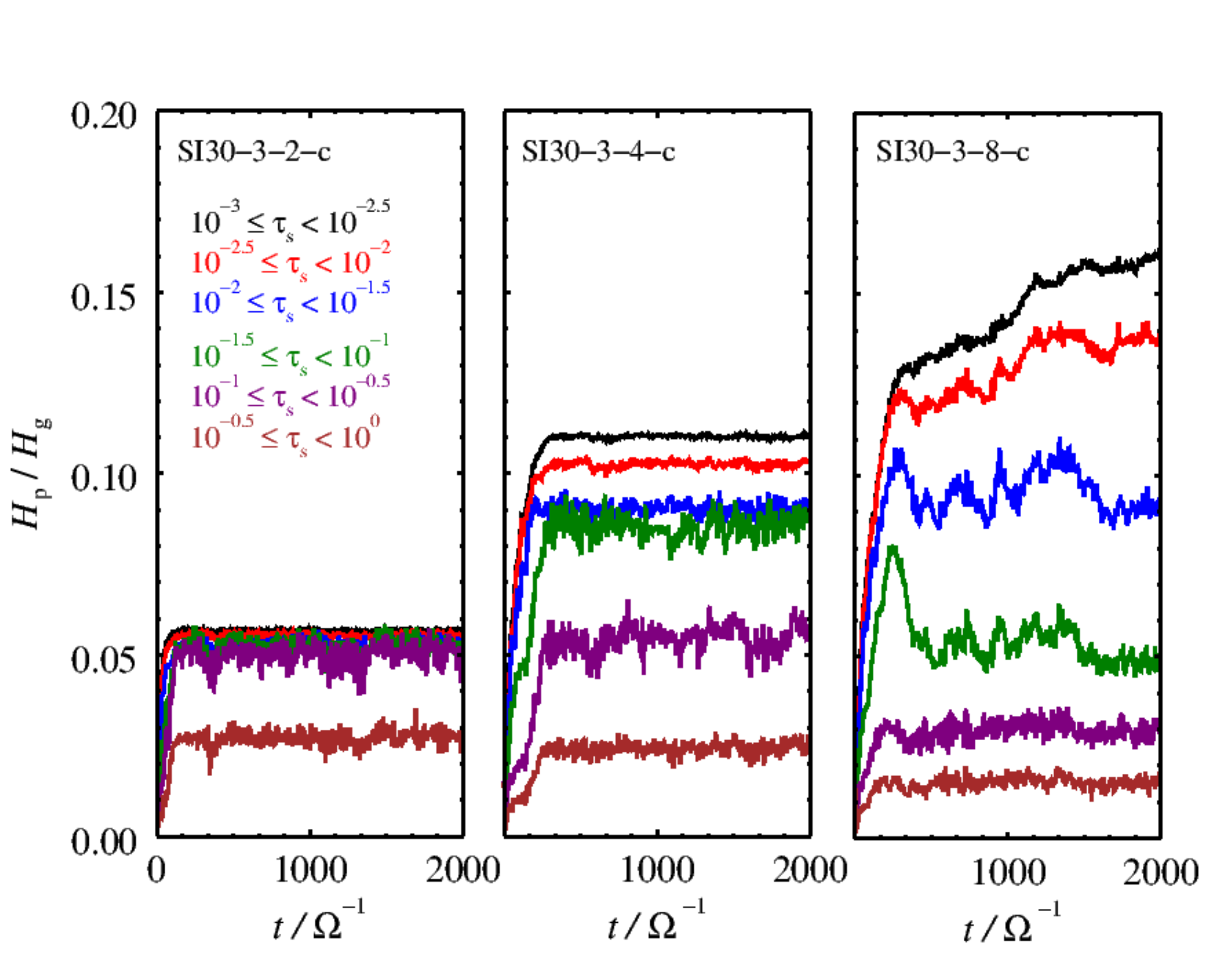}
  \caption{Particle scale height evolution of run SI30-3-2-c, run SI30-3-4-c and run SI30-3-8-c. The panels from left to right show the evolution of the system in simulation boxes of $0.2 H_{\rm{g}} \times 0.2 H_{\rm{g}}$, $0.4 H_{\rm{g}} \times 0.4 H_{\rm{g}}$ and $0.8 H_{\rm{g}} \times 0.8 H_{\rm{g}}$, respectively. The largest particles converge to the same height in the first two cases, but particles with $\tau_{\rm{s}} < 10^{-1}$ are diffused to heights larger by about a factor of two. The plot on the right, where the simulation box size is the largest, shows unusual non-convergent behavior due to the development of shock-like patterns in the radial component of the gas velocity (discussed in \ref{Appendix1}). \label{hp:7}}
\end{minipage}\hfill
\begin{minipage}[t]{1\columnwidth}
  \centering
  \includegraphics[width=\columnwidth]{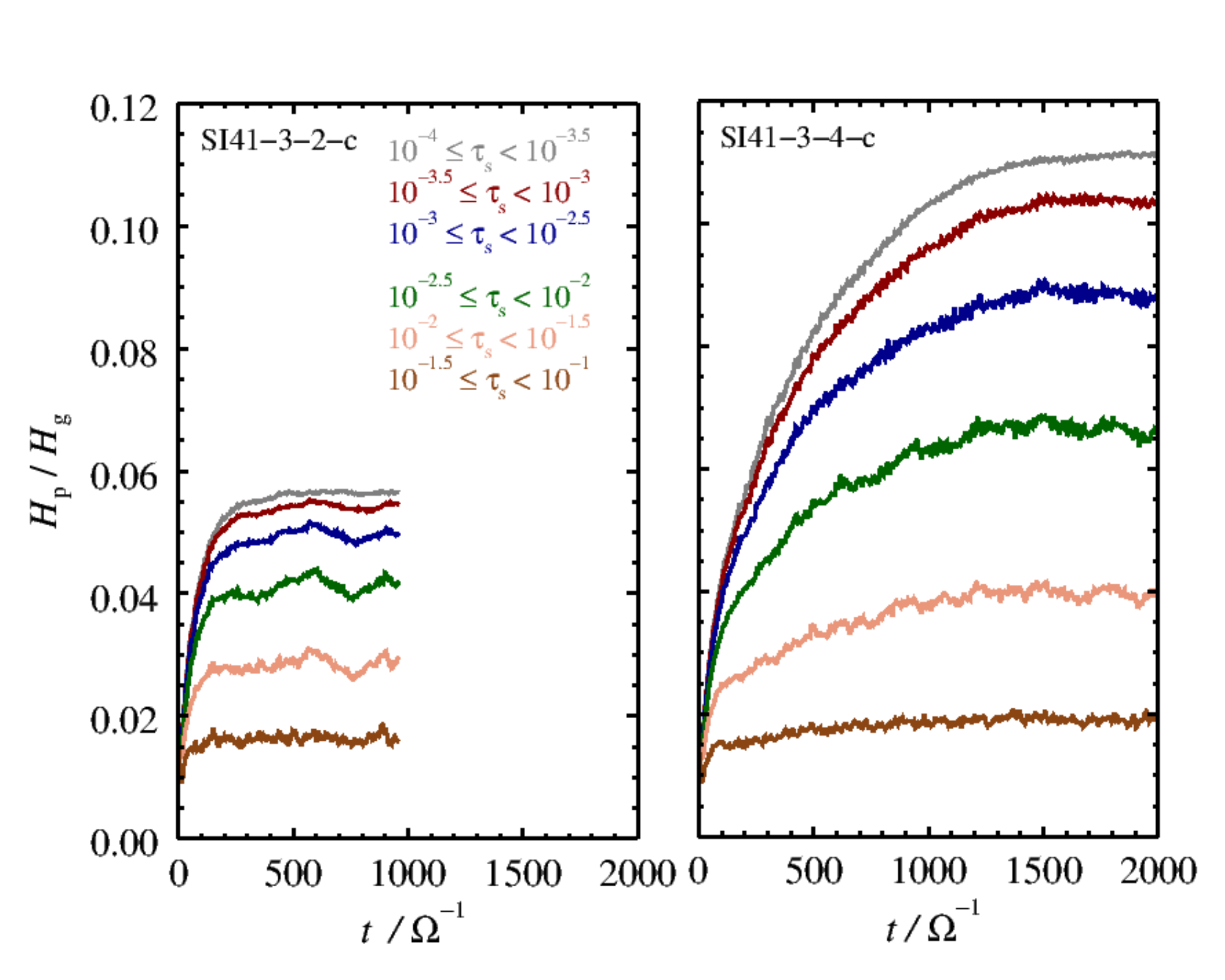}
  \caption{Particle scale height evolution of runs SI41-3-2-c and SI41-3-4-c. The particle scale height of the small species increases by a factor of two when changing the dimensions from $0.2 H_{\rm{g}} \times 0.2 H_{\rm{g}}$ in the left panel to $0.4 H_{\rm{g}} \times 0.4 H_{\rm{g}}$ in the right panel, while that of the largest particles converges to about the same height. \label{hp:8}}
\end{minipage}\hfill
\end{figure*}

Figure \ref{hp:1} and Fig. \ref{hp:2} show the evolution of the particle scale height in the case of discrete steep particle size distribution. Figure \ref{hp:1} shows the system where the participating particles are $\tau_{\rm{s}} = 10^{-3} - 10^0$, with each particle species differentiated by a unique color. The system is first modeled in a $0.2 H_{\rm{g}} \times 0.2 H_{\rm{g}}$ box, shown on the left. Each particle species saturates to a different scale height in such a way that $H_{\rm{p}}$ increases with decreasing Stokes number. After saturation, the scale heights show some fluctuations, especially for the smaller species. We ran the same model in a larger box of  $0.4 H_{\rm{g}} \times 0.4 H_{\rm{g}}$, and the result is shown on the right panel in Fig. \ref{hp:1}. This was done to check that the turbulent diffusion of the smaller solids is not limited by the size of the simulation box. The saturated scale height of the smallest species is indeed larger for the larger box. This indicates that the vertical box size prevented these solids from reaching the scale heights as they should. On the other hand, the larger species saturate to about the same scale height, indicating convergence.

We compared this to the case with the smaller participating particle species, namely $\tau_{\rm{s}} = 10^{-4} - 10^{-1}$. The result is shown in Fig. \ref{hp:2}. Given that the particles are smaller, it takes longer for the system to settle into equilibrium. This is due to the larger settling time \citep{Nakagawa1986, Dubrulle1995, Youdin2007a}

\begin{equation}
t_{\rm{settle}} = \frac{1}{\tau_{\rm{s}} \varOmega}.
\end{equation}

\noindent  As above, we increased the box size to ensure that the vertical diffusion of the particles is not limited by the height of the simulation domain. The panel in the middle of Fig. \ref{hp:2} shows that the scale height of the smallest species increased and that of the larger species decreased compared to the case of the smaller box. In addition to increasing the box size, we also conducted another convergence test, where we increased the total number of particles by a factor of four. The result is shown in the right-most panel of Fig. \ref{hp:2}. None of the species settle to a constant scale height within approximately 300 orbits.

Next, we considered the case of shallow and discrete particle size distribution with a power-law index of $q = 3$. In other words, the system now has more mass in larger particle species. The evolution of both SI30-3-2-d and SI41-3-2-d is shown in Fig. \ref{hp:3} and Fig. \ref{hp:4}. Comparing the corresponding panel in Fig. \ref{hp:1} and Fig. \ref{hp:2}, we see that the particles in the case of both $q = 3$ and $q = 4$ reach similar scale heights. The particle scale heights in Fig. \ref{hp:4} and the first panel in Fig. \ref{hp:2} are similar as well. This implies that the level of turbulence is similar both when the particle distribution is steep and shallow, which is consistent with the similar dispersions of the radial drift velocity of particles in the respective models found in Sect. \ref{SectionDiscreteVelocity}.

\subsection{Continuous size distribution}

We turn to studying the evolution of the continuously distributed particle sizes. Since in this case each particle has a unique size, we group them into six bins for illustration purposes in the following discussions.

In Fig. \ref{hp:5} and Fig. \ref{hp:6}, we show the scale height evolution of runs SI30-4-2-c and SI41-4-2-c. Similar to the previously presented cases, where the particles are distributed in a discrete way in terms of their size, the large species remain closer to the midplane, while the self-generated turbulence diffuses small particles to larger heights. Once again, we conducted convergence tests in terms of simulation box size. As shown in Fig. \ref{hp:5}, in the case of size range $\tau_{\rm{s}} = 10^{-3} - 10^{0}$, the smallest species saturate to a larger $H_{\rm{p}}$ in the $0.4 H_{\rm{g}} \times 0.4 H_{\rm{g}}$ box than in the smaller box, while the larger solids converge to the same scale heights. Figure \ref{hp:6} shows the evolution of the particle scale height, when the particles are $\tau_{\rm{s}} = 10^{-4} - 10^{-1}$. The particle species do not settle to a constant height during the simulation time in either the $0.2 H_{\rm{g}} \times 0.2 H_{\rm{g}}$ box, nor the one double that size, although the smaller species reach large heights in the latter case. 

Finally, in Fig. \ref{hp:7} and Fig. \ref{hp:8} we show the evolution of the system with shallow continuous particle size distribution. Figure \ref{hp:7} shows the scale height evolution of runs SI30-3-2-c, SI30-3-4-c and SI30-3-8-c, which were performed in boxes of $0.2 H_{\rm{g}} \times 0.2 H_{\rm{g}}$, $0.4 H_{\rm{g}} \times 0.4 H_{\rm{g}}$ and $0.8 H_{\rm{g}} \times 0.8 H_{\rm{g}}$, respectively. We see that in run SI30-3-4-c the four smallest particle bins reach heights almost a factor of two larger compared to the run in the smaller box. To test the convergence of the system, we increased the box size again, to $0.8 H_{\rm{g}} \times 0.8 H_{\rm{g}}$. The scale heights of the three largest species reach a lower scale height than before, while the smaller species are diffused to larger heights. This unusual behavior is accompanied by the formation of shock-like patterns in the radial component of the gas velocity. We discuss this in more detail in Appendix \ref{Appendix1}. 
Figure \ref{hp:8} shows the evolution of the same model with smaller particles. Since all species settle to a relatively constant height, we stop the run after about 140 orbits. However, when doubling the domain size to $0.4 H_{\rm{g}} \times 0.4 H_{\rm{g}}$ the smallest species bins saturate at heights almost twice as large compared to the smaller box.

Comparing Fig. \ref{hp:7} and Fig. \ref{hp:8} with Fig. \ref{hp:5} and Fig. \ref{hp:6}, respectively, we see that in the latter case where the size distribution is steep ($q = 4$), all species reach lower scale heights in general. The same trend is present once we increase the box size from our default size. This is likely due to the difference in the number density of the solids. By increasing the mass in larger particle species, we increase the availability of particles that actively participate in the streaming instability which generates turbulence. Thus, particles in runs with continuous and shallow size distribution experience more diffusion and the particle layers become more puffed up around the midplane.

\subsection{Effect of multiple particle species}

\begin{figure*}[t!]
\centering
\begin{minipage}[t]{1\columnwidth}
\centering
\includegraphics[width=1\columnwidth]{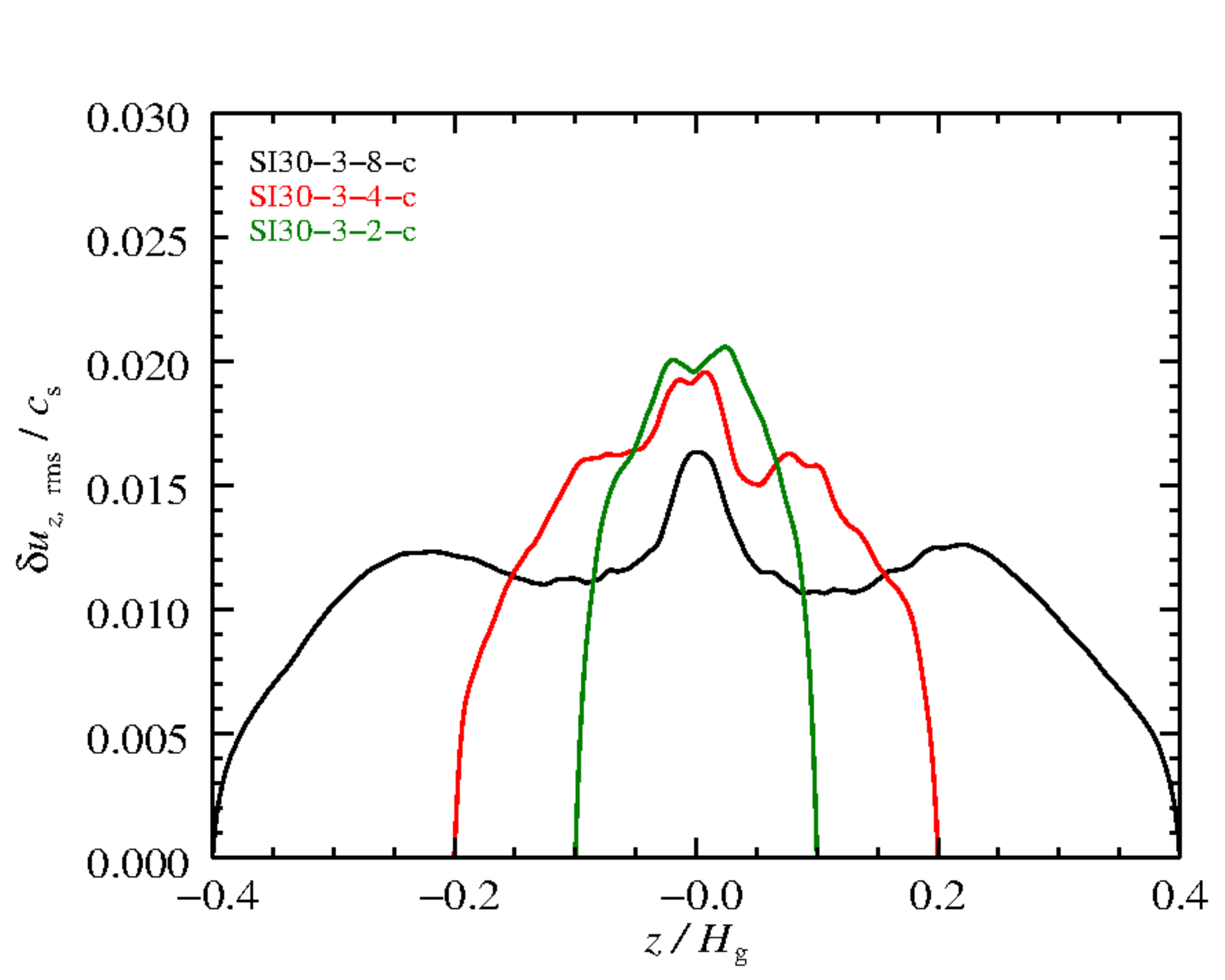}
\caption{Vertical gas velocity fluctuation with respect to height in the case of runs SI30-3-8-c, SI30-3-4-c and SI30-3-2-c averaged over approximately 15 orbits starting at $t \approx 1000 \varOmega^{-1}$. The relatively uniform level of fluctuation further away from the midplane implies that the turbulence is not limited to the region close to the midplane ($z = 0$), where the large particles reside. The turbulence extends to larger scale heights, thus affecting the small particle species that reside there.}
\label{GasVelocityFluct}
\end{minipage}\hfill
\begin{minipage}[t]{1\columnwidth}
\centering
\includegraphics[width=1\columnwidth]{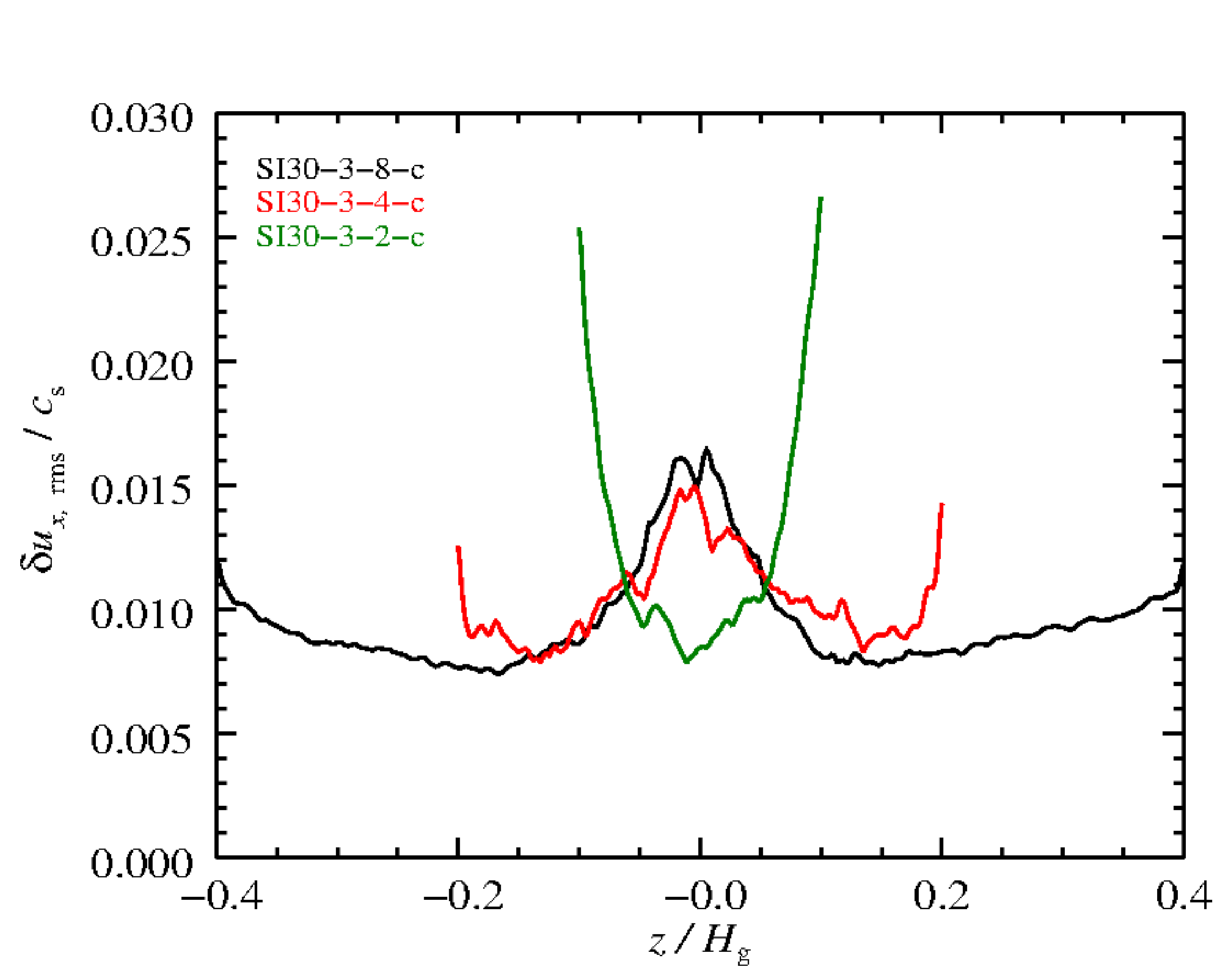}
\caption{Horizontal gas velocity fluctuation with respect to height in the case of runs SI30-3-8-c, SI30-3-4-c and SI30-3-2-c averaged over approximately 15 orbits starting at $t \approx 1000 \varOmega^{-1}$. The level of fluctuation peaks around the midplane and is relatively uniform throughout the rest of the domain for the runs in the $0.4 H_{\rm{g}} \times 0.4 H_{\rm{g}}$ and $0.8 H_{\rm{g}} \times 0.8 H_{\rm{g}}$ boxes, while in the smallest box $ \delta u_{x,\rm{rms}}$ varies by more than a factor of two.}
\label{GasVelocityFluctHor}
\end{minipage}\hfill
\begin{minipage}[t]{1\columnwidth}
\centering
\includegraphics[width=1\columnwidth]{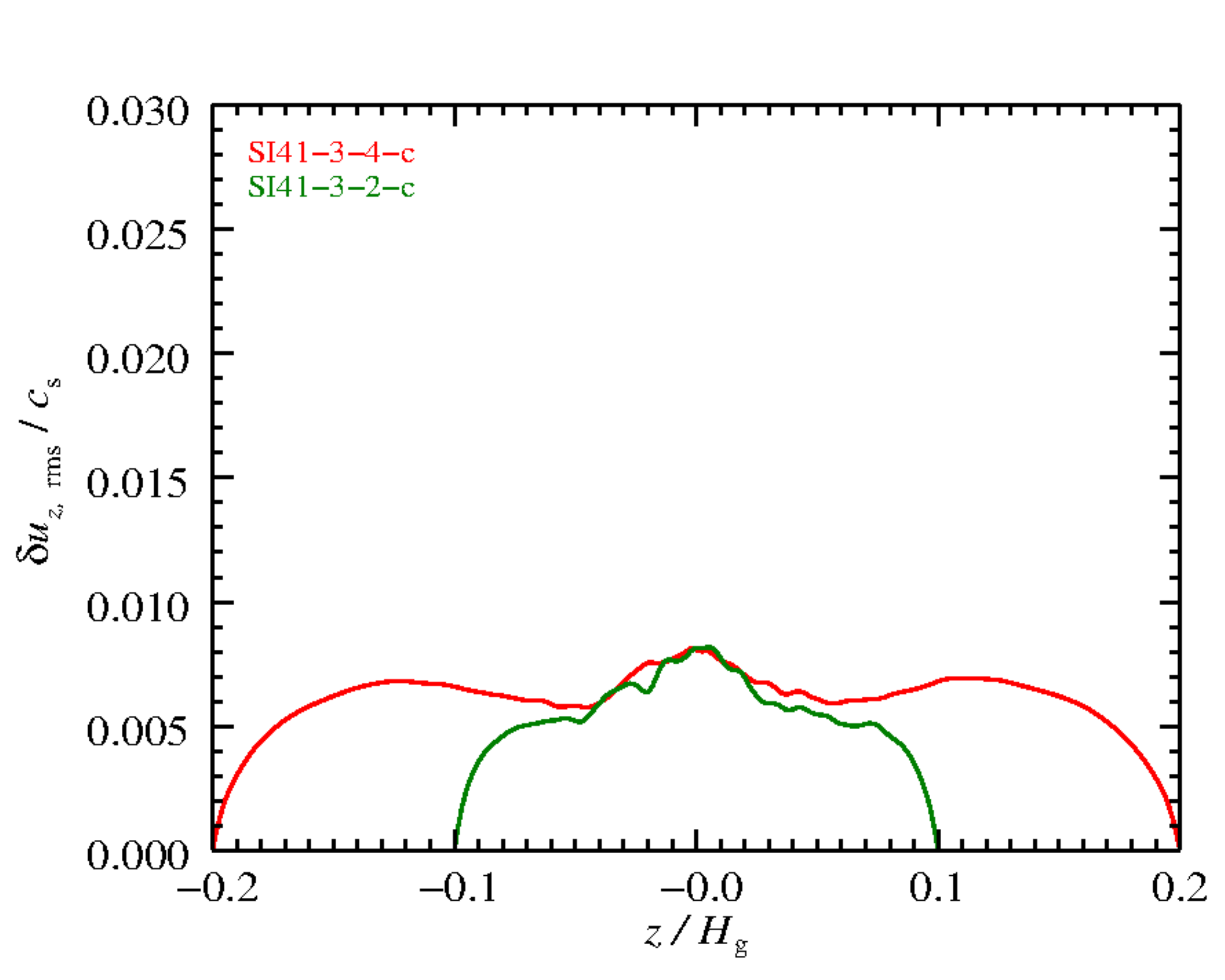}
\caption{Vertical gas velocity fluctuation with respect to height in the case of runs SI41-3-4-c and SI41-3-2-c averaged over approximately 15 orbits starting at $t \approx 500 \varOmega^{-1}$. The level of fluctuation is relatively uniform for the two runs both in terms of height and box size.}
\label{GasVelocityFluct41}
\end{minipage}\hfill
\begin{minipage}[t]{1\columnwidth}
\centering
\includegraphics[width=1\columnwidth]{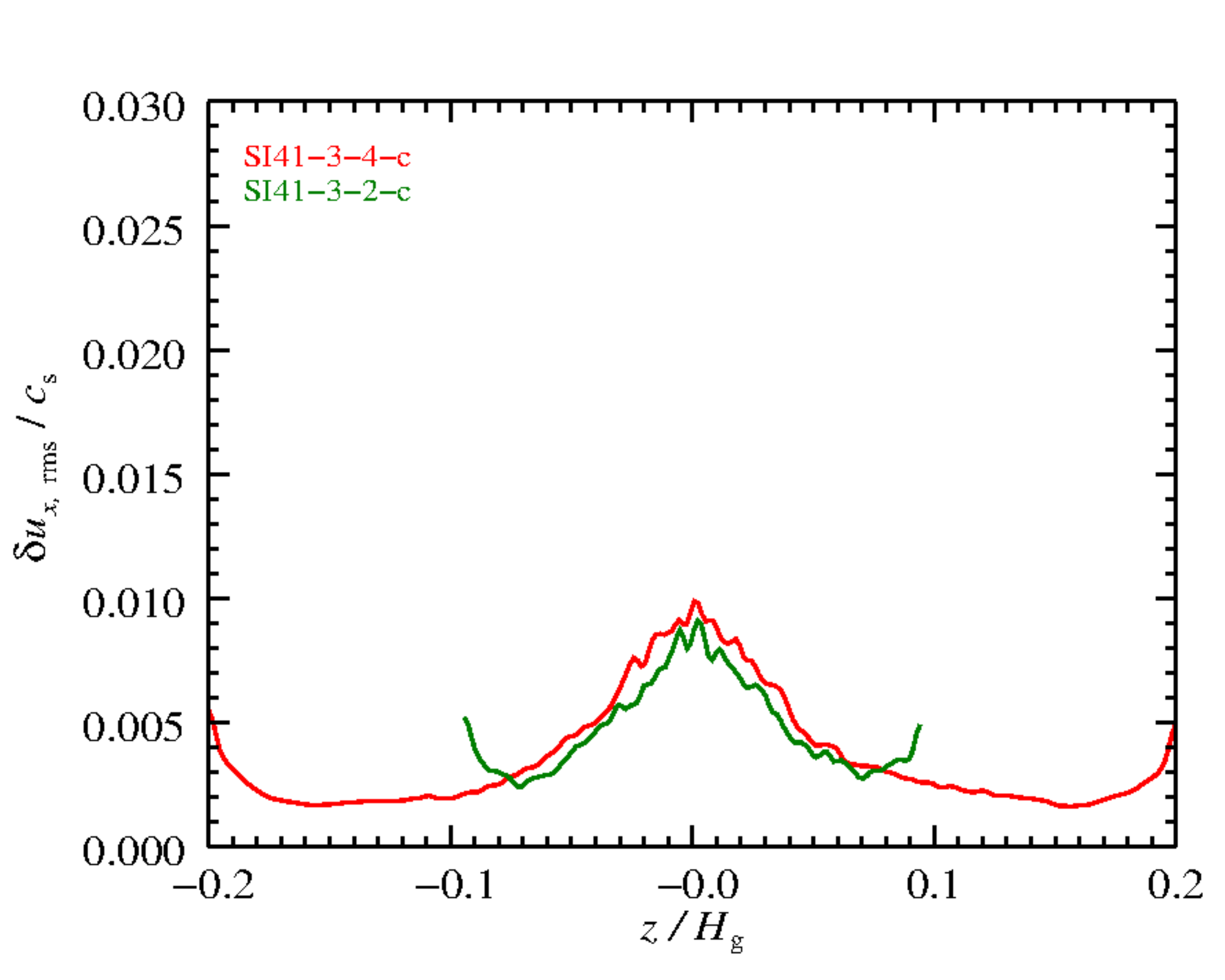}
\caption{Horizontal gas velocity fluctuation with respect to height in the case of runs SI41-3-4-c and SI41-3-2-c averaged over approximately 15 orbits starting at $t \approx 500 \varOmega^{-1}$. The level of fluctuation peaks around the midplane for both box sizes and is relatively uniform throughout the rest of the domain.}
\label{GasVelocityFluctHor41}
\end{minipage}\hfill
\end{figure*}

The effect of multiple particle sizes in the same system is evident from the change in mean radial drift velocity relative to the single species case and the effect of the self-regulated diffusion driven by the larger species. Compared to single-species simulations, there is also significant difference in terms of the scale height each particle species saturates to. \cite{Carrera2015} conducted single-species streaming instability simulations of a variety of Stokes numbers. They found that particles of $10^{-3} \le \tau_{\rm{s}} \le 10^{0}$ saturate to $H_{\rm{p}} \approx 10^{-2} H_{\rm{g}}$ (see also \citealt{Yang2014, Yang2017}). In general, we see that most particle species, especially those with $\tau_{\rm{s}} \le 10^{-0.5}$, are diffused to heights much larger than that. This is the consequence of the strong turbulence driven by the large particles. The vertical velocity fluctuations in the gas are a measure of the level of turbulence driven by the particles. Figure \ref{GasVelocityFluct} shows the root-mean-square of the vertical gas velocity in terms of height for the runs in three different box sizes, namely run SI30-3-2-c, SI30-3-4-c and SI30-3-8-c. The level of turbulence is not only significant near the midplane ($z = 0$), but also at larger heights. Thus, the smaller particle species which reside at larger heights experience the effect of the stirring from the large particles near the midplane.  
In Fig. \ref{GasVelocityFluctHor} we show the horizontal component of the gas velocity fluctuations as a function of height. The curves in the case of runs SI30-3-4-c and SI30-3-8-c peak around the midplane and drop to an approximately uniform value beyond $\pm 0.1 H_{\rm{g}}$ until the increase near the walls. In the case of run SI30-3-2-c, the fluctuation is at a minimum around the midplane and increases towards the walls of the simulation domain.

In Fig. \ref{GasVelocityFluct41} and Fig. \ref{GasVelocityFluctHor41} we show the vertical and horizontal velocity fluctuations of runs SI41-3-2-c and SI41-3-4-c. Both $\delta u_{z,\rm{rms}}$ and $\delta u_{x,\rm{rms}}$ show similar behavior in terms of box size. The vertical component shows uniformity with respect to height, while the horizontal component peaks near the midplane and decreases towards the walls.

\section{Radial and vertical diffusion}

\label{DiffusionSection}

The mutual drag between the solid and gas components triggers the streaming instability and generates turbulence. This self-regulated turbulence in return stirs up the particles which experience diffusion in both the radial and vertical dimensions. To understand the influence of each particle species on the level of turbulence and their response to it, we measured both the radial and vertical diffusion coefficients for each particle size. We did this only for the runs where the particles were distributed in a discrete fashion in terms of their size. The continuous runs cannot be used to measure the turbulent diffusion, since in that case, every particle has a unique size and calculating the diffusion of each species with respect to the rest would be meaningless due to the differential radial drift.

The radial diffusion coefficient, $D_{x}$, represents the degree of random walk each species experiences due to the self-regulated turbulence.
It is calculated following \cite{Johansen2007} as

\begin{equation}
D_x = \frac{1}{2} \frac{\rm{d} \sigma_{\it{x}}^2}{\rm{d}\it{t}}.
\label{Eq19}
\end{equation}

\begin{figure*}[th!]
\centering
\begin{minipage}[t]{\columnwidth}
\centering
\includegraphics[width=1\columnwidth]{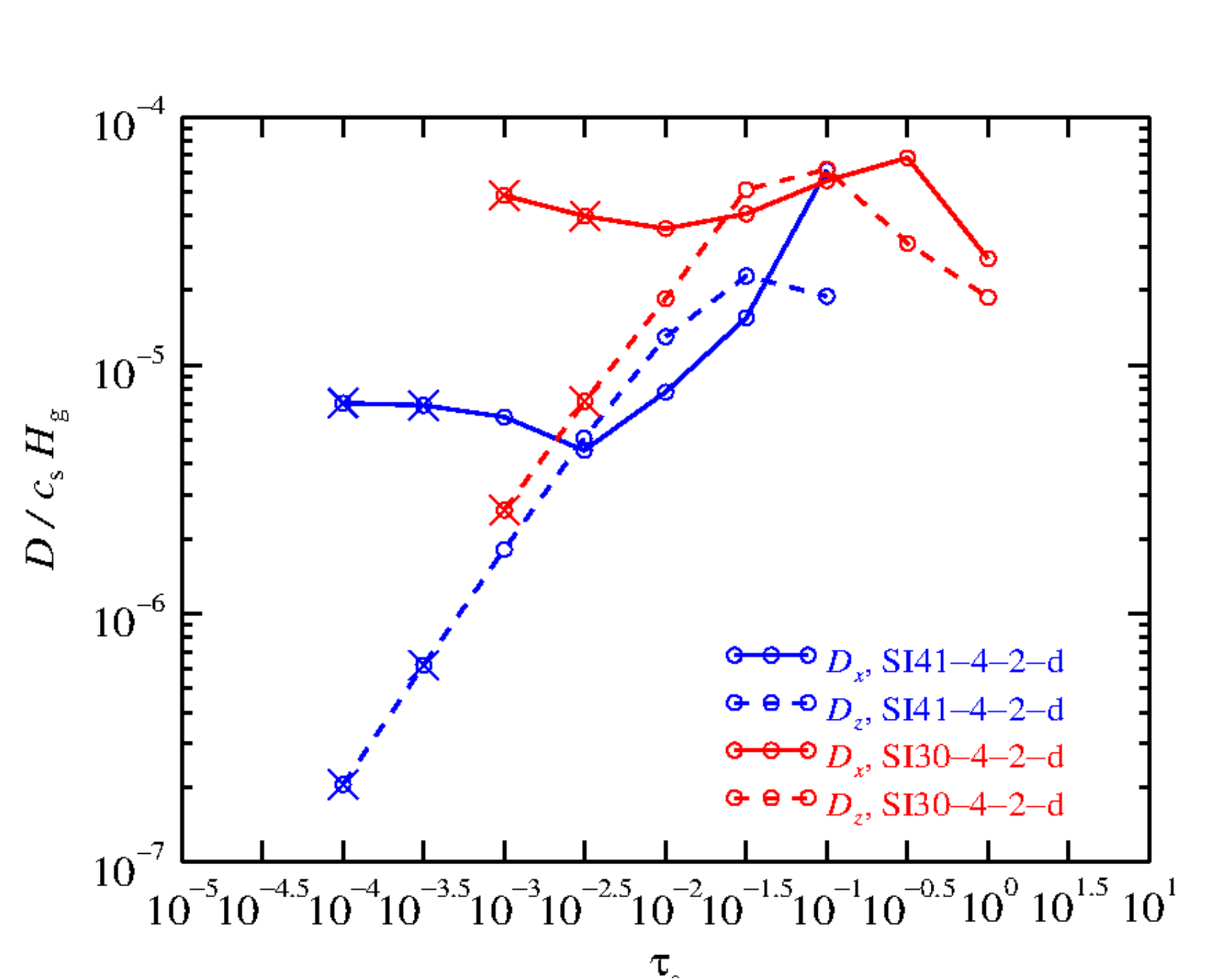}
\caption{Radial and vertical diffusion coefficients with respect to Stokes number for runs SI41-4-2-d and SI30-4-2-d. Here, $D_{x}$ is approximately constant for small Stokes numbers and increases with increasing particle size as $\tau_{\rm{s}} \ge 10^{-2.5}$ in the case of run SI41-4-2-d. The vertical diffusion coefficient increases with particle size until $\tau_{\rm{s}} \approx 10^{-1.5}$. Above this value it decreases as the particles are less coupled to the gas. Both the radial and vertical diffusion coefficients are higher in the case of run SI30-4-2-d, since  this model is more abundant in solids with $\tau_{\rm{s}} \gtrsim 10^{-2}$ which are responsible for triggering the streaming instability. The diffusion coefficients corresponding to species that saturate to scale heights similar to the maximum allowed scale height in the box are marked with a cross.}
\label{Diffusion4d2}
\end{minipage}\hfill
\begin{minipage}[t]{\columnwidth}
\centering
\includegraphics[width=1\columnwidth]{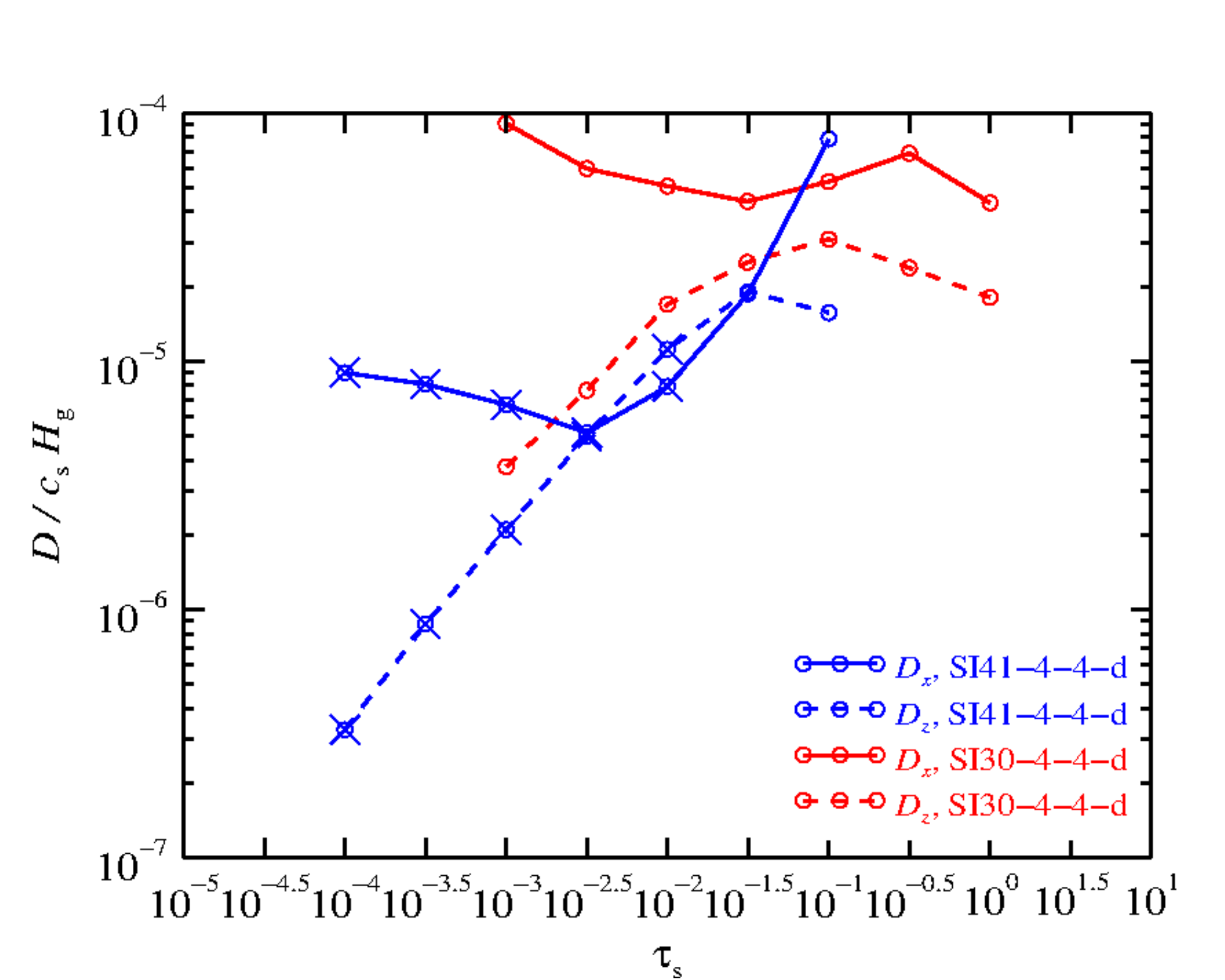}
\caption{Radial and vertical diffusion coefficients with respect to Stokes number for runs SI41-4-4-d and SI30-4-4-d. We doubled the box size to $0.4 H_{\rm{g}} \times 0.4 H_{\rm{g}}$ compared to the simulations in Fig. \ref{Diffusion4d2}. Compared to the smaller box, both $D_{x}$ and $D_{z}$ converge to approximately the same values for each Stokes number. The exception seen in curve $D_{z, 30}$ at intermediate Stokes numbers is likely a consequence of the fluctuations seen in their particle scale height in the smaller box (see Fig. \ref{hp:1}). The diffusion coefficients corresponding to species that do not saturate within the simulation time are marked with a cross.}
\label{Diffusion44d2}
\end{minipage}
\end{figure*}
\noindent Here, $\sigma_x^2$ is the variance of the radial displacement that each species covers at discrete time intervals. We applied linear regression on the resulting curve to find the radial diffusion coefficient of each species.

The vertical diffusion coefficient, $D_z$, is calculated using the scale height of the particles, such that \citep{Dubrulle1995, Youdin2007a}

\begin{equation}
\frac{H_{\rm{p}}}{H_{\rm{g}}} = \sqrt{\frac{\delta}{\tau_{\rm{s}} + \delta}}.
\label{Eq20}
\end{equation}

\noindent Here, $\delta$ is a dimensionless measure of the diffusion coefficient, such that $D_z = \delta c_{\rm{s}} H_{\rm{g}}$ \citep{Johansen2005, Johansen2014}.

\begin{figure*}[ht!]
\centering
\begin{minipage}[t]{\columnwidth}
\includegraphics[width=1\columnwidth]{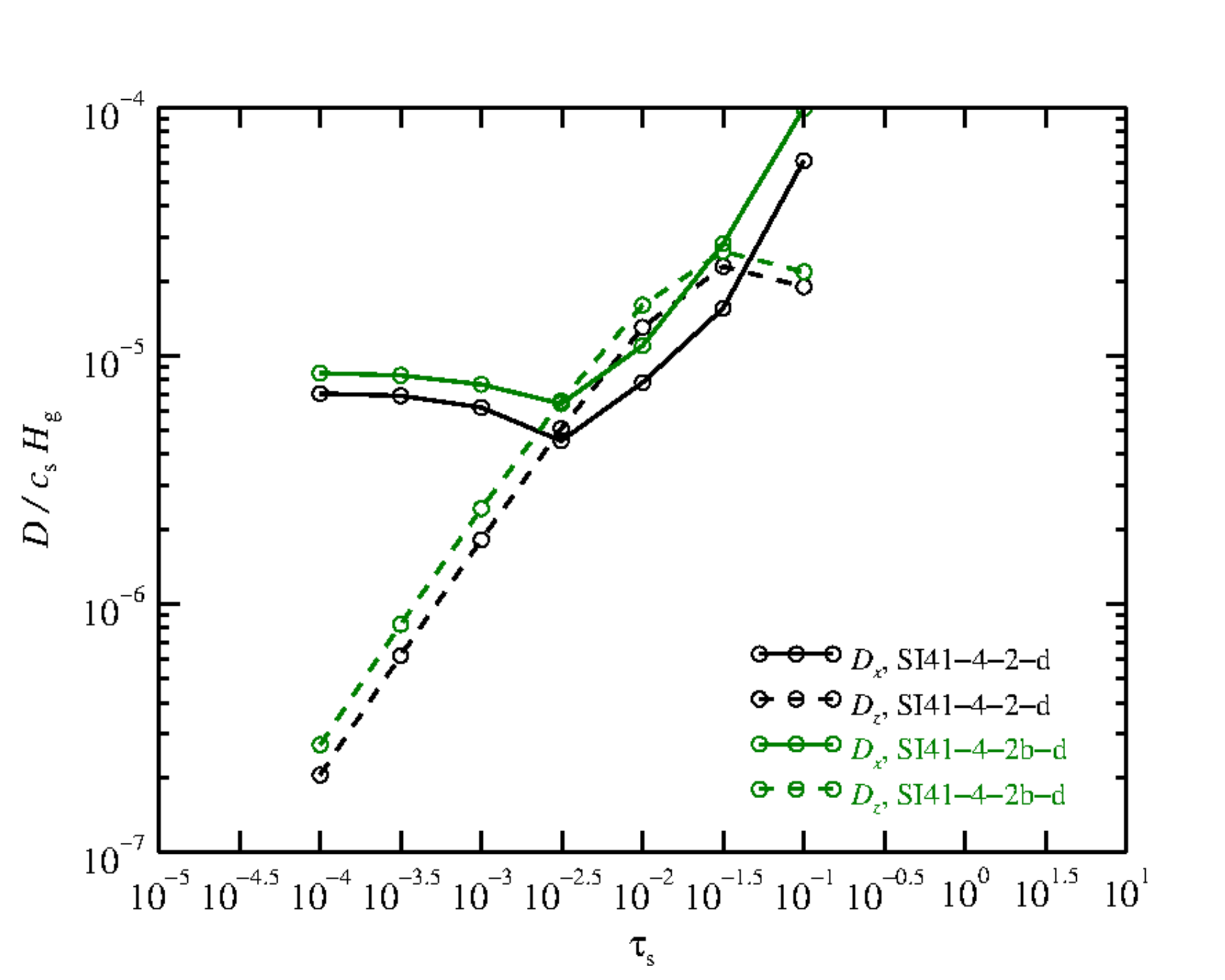}
\caption{Radial and vertical diffusion coefficients with respect to Stokes number for runs SI41-4-2-d and SI41-4-2b-d. Both $D_{x}$ (solid curves) and $D_{z}$ (dashed curves) show relatively small deviation, suggesting that numerical resolution does not appreciably influence the level of turbulence felt by the particles.}
\label{Diffusion4Conv}
\end{minipage}
\hfill
\begin{minipage}[t]{\columnwidth}
\centering
\includegraphics[width=1\columnwidth]{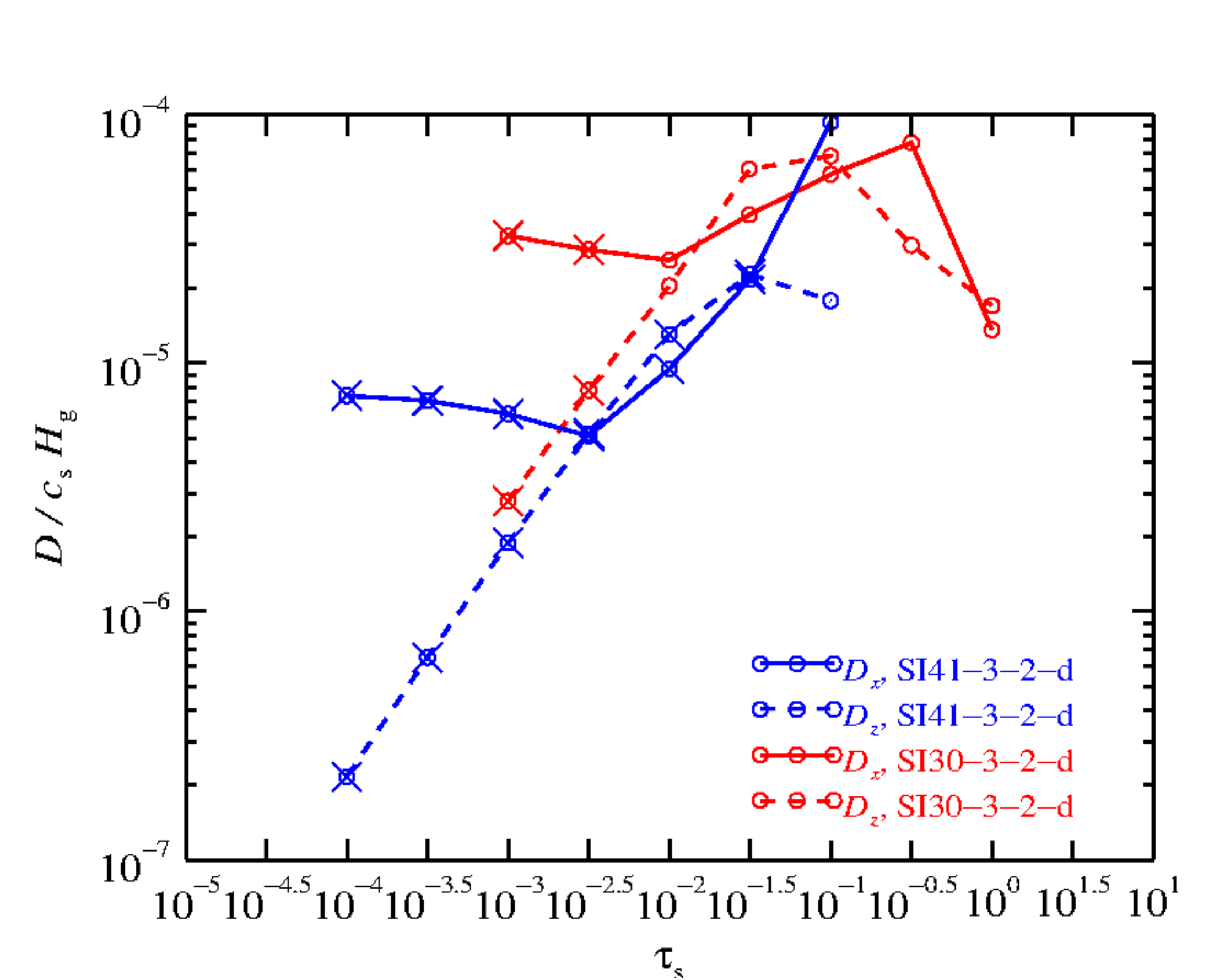}
\caption{Radial and vertical diffusion coefficients with respect to Stokes number for runs SI41-3-2-d and SI30-3-2-d. Both $D_x$ and $D_z$ take on similar values as in Fig. \ref{Diffusion4d2}, where the particle size distribution is shallow ($q = 4$). The diffusion coefficients corresponding to species that saturate to scale heights similar to the maximum allowed scale height in the box or do not saturate within the simulation time are marked with a cross.} 
\label{Diffusion3d2}
\end{minipage}
\end{figure*}

\subsection{Steep size distribution}
We first considered the case of steep particle size distribution, with a power-law index of $q = 4$.
In Fig. \ref{Diffusion4d2} we present the relationship between diffusion coefficient and Stokes number and compare the dependence of $D_{x}$ and $D_{z}$ on particle size. As seen in Fig. \ref{Diffusion4d2}, $D_{z}$ increases with increasing Stokes number in both run SI41-4-2-d (blue dashed curve) and SI30-4-2-d (red dashed curve) for $\tau_{\rm{s}} \le 10^{-1.5}$. This trend arises, since as we discussed in Sect. \ref{SectionScaleHeight}, smaller particles saturate to larger scale heights. As a consequence, smaller species can reside in regions where the level of turbulence is somewhat lower than near the midplane, as seen in Fig. \ref{GasVelocityFluct}. The decreasing diffusion coefficients of the smaller species could also be influenced by the box size, such that the random walk of the particles close to the domain walls is restricted. On Fig. \ref{Diffusion4d2} we marked these values with crosses. When $\tau_{\rm{s}} \ge 10^{-1.5}$, the vertical diffusion coefficient decreases with increasing Stokes number. These species saturate at small scale heights and hence feel the full strength of the turbulent diffusion in the midplane layer.

The radial diffusion coefficient, $D_{x}$, for run SI30-4-2-d and SI41-4-2-d is relatively constant for $\tau_{\rm{s}} \le 10^{-2}$, since as Figure \ref{hp:1} and Fig. \ref{hp:2} show, particles of these sizes saturate to similar scale heights. Above $\tau_{\rm{s}} \approx 10^{-2.5}$, $D_x$ increases with increasing Stokes number, since in contrast with the smaller species, they reside closer to the midplane, where they can experience more turbulence. This behavior can be seen in both Fig. \ref{GasVelocityFluctHor} and Fig. \ref{GasVelocityFluctHor41}.
Comparing the curves that correspond to run SI41-4-2-d (blue curves) and SI30-4-2-d (red curves), we see that for all particle species both the vertical and radial coefficients are higher in the latter case. This is the consequence of the presence of more particles with $\tau_{\rm{s}} \gtrsim 10^{-2}$, which makes it easier for the streaming instability to operate and drive the turbulence \citep{Bai2010a}. In other words, the larger the abundance of the actively participating particles is, the stronger the streaming turbulence becomes.

As shown in Fig. \ref{hp:2}, the vertical diffusion of the smallest particle species is limited by the vertical domain, since their scale height increases as we double the box size. We measure $D_x$ and $D_{z}$ in the $0.4 H_{\rm{g}} \times 0.4 H_{\rm{g}}$ box, and show the result in Fig. \ref{Diffusion44d2}. Compared to the case of the smaller simulation box shown in Fig. \ref{Diffusion4d2}, there is relatively little change in the radial diffusion coefficients for both runs, which implies convergence. The vertical diffusion coefficient $D_{z, 41}$ (blue dashed curves) in both Fig. \ref{Diffusion4d2} and Fig. \ref{Diffusion44d2} take on similar values for each particle species. On the other hand, $D_{z, 30}$ (red dashed curves) takes on slightly different values in both boxes, especially for $10^{-2} < \tau_{\rm{s}} < 10^{-0.5}$. This could also be a consequence of the fluctuations of the particle scale height in the smaller box, as see in Fig. \ref{hp:1}. We marked the diffusion coefficients that correspond to species which do not saturate within the simulation time with crosses.

In Fig. \ref{Diffusion4Conv}, we present the result of a numerical convergence test performed on run SI41-4-2-d. We increased the number of grid cells in the radial and vertical dimensions from our default value of $128 \times 128$ to $256 \times 256$, while at the same time leaving the physical box size unchanged, such that $L_x = L_z = 0.2 H_{\rm{g}}$. Figure \ref{Diffusion4Conv} shows convergence may have been reached against numerical resolution.

\subsection{Shallow size distribution}

We also measured the diffusion coefficients in the case of shallow particle size distribution.
Figure \ref{Diffusion3d2} shows $D_x$ and $D_z$ of runs SI41-3-2-d and SI30-3-2-d. Comparing the diffusion coefficients to the ones in Fig. \ref{Diffusion4d2}, we see that there is not a significant increase in either $D_x$, nor $D_z$. Both Fig. \ref{hp:3} and Fig. \ref{hp:4} show the fluctuations of the equilibrium scale heights, which according to Equation \eqref{Eq20}, are responsible for setting $D_z$. The diffusion coefficients which correspond to species that either saturate to scale heights similar to the maximum allowed scale height in the box or do not saturate within the simulation time are marked with crosses.

\section{Implications}

\label{SectionImplication}

Particle scale height is a key parameter in grain growth by coagulation \citep{Zsom2011, Drazkowska2013}. As shown in Sect. \ref{SectionScaleHeight} and Sect. \ref{DiffusionSection}, every particle species saturates to a unique scale height and thus has a response to the self-regulated turbulence. Since coagulation is driven by the relative velocity of the solid materials as well as their local number density, its efficiency is strongly influenced by the equilibrium particle scale height of the given particle species.

We apply here the main results presented in the paper to protoplanetary disks by showing the evolution of particle radial position, $r$, as a function of time, $t$, taking into account both radial drift and diffusion over approximately $300$ orbits.

\begin{figure}[th!]
\centering
\includegraphics[width=1\columnwidth]{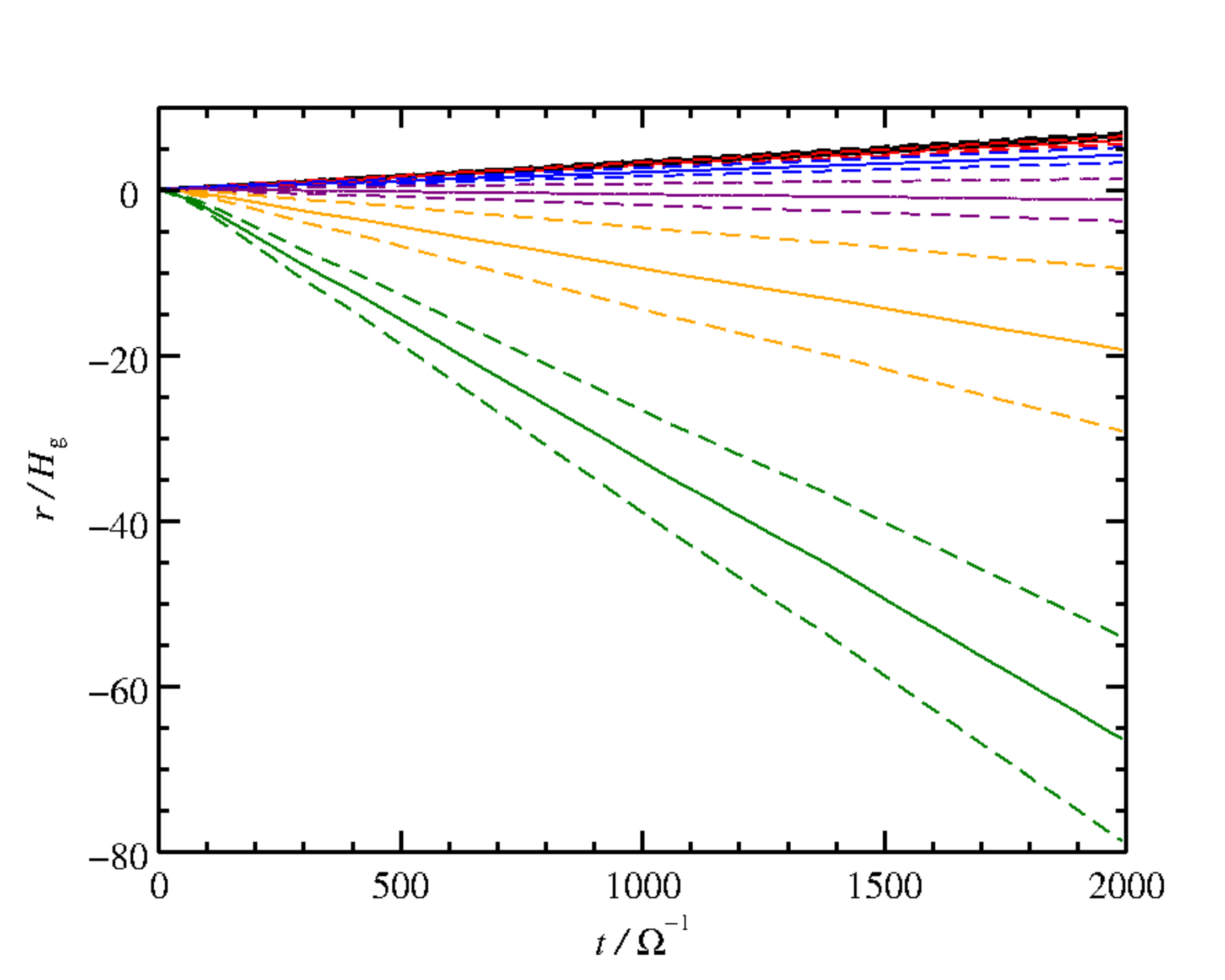}\llap{\makebox[0.85\columnwidth][l]{\raisebox{1.5cm}{\includegraphics[height=2cm]{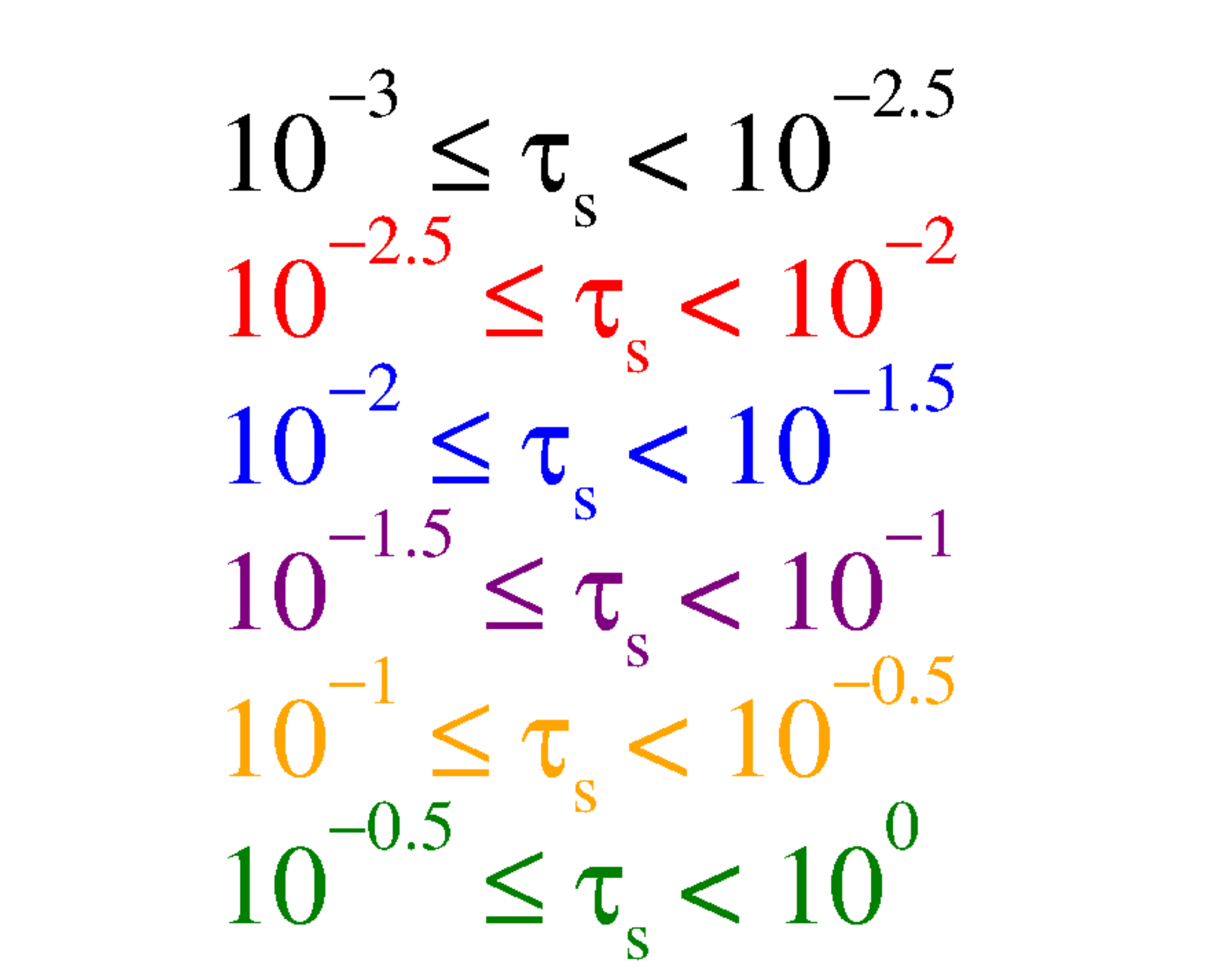}}}}
\caption{Evolution of particle radial positions with respect to time in the case of run SI30-3-2-c. Since the particles are distributed continuously in terms of their size, we grouped them into six particle size bins. The mean position of the particles in a given size range is marked by the solid curves, while the dashed curves mark the spread of particle positions by differential radial drift and diffusion. Particles with $\tau_{\rm{s}} \ge 10^{-1}$ drift and diffuse inwards, while particles of $\tau_{\rm{s}} \le 10^{-1.5}$ tend to move outwards in the disk. The spreading of each particle species is in fact dominated by the differential radial drift instead of diffusion.}
\label{rvst}
\end{figure}

\begin{figure}[th!]
\centering
\includegraphics[width=1\columnwidth]{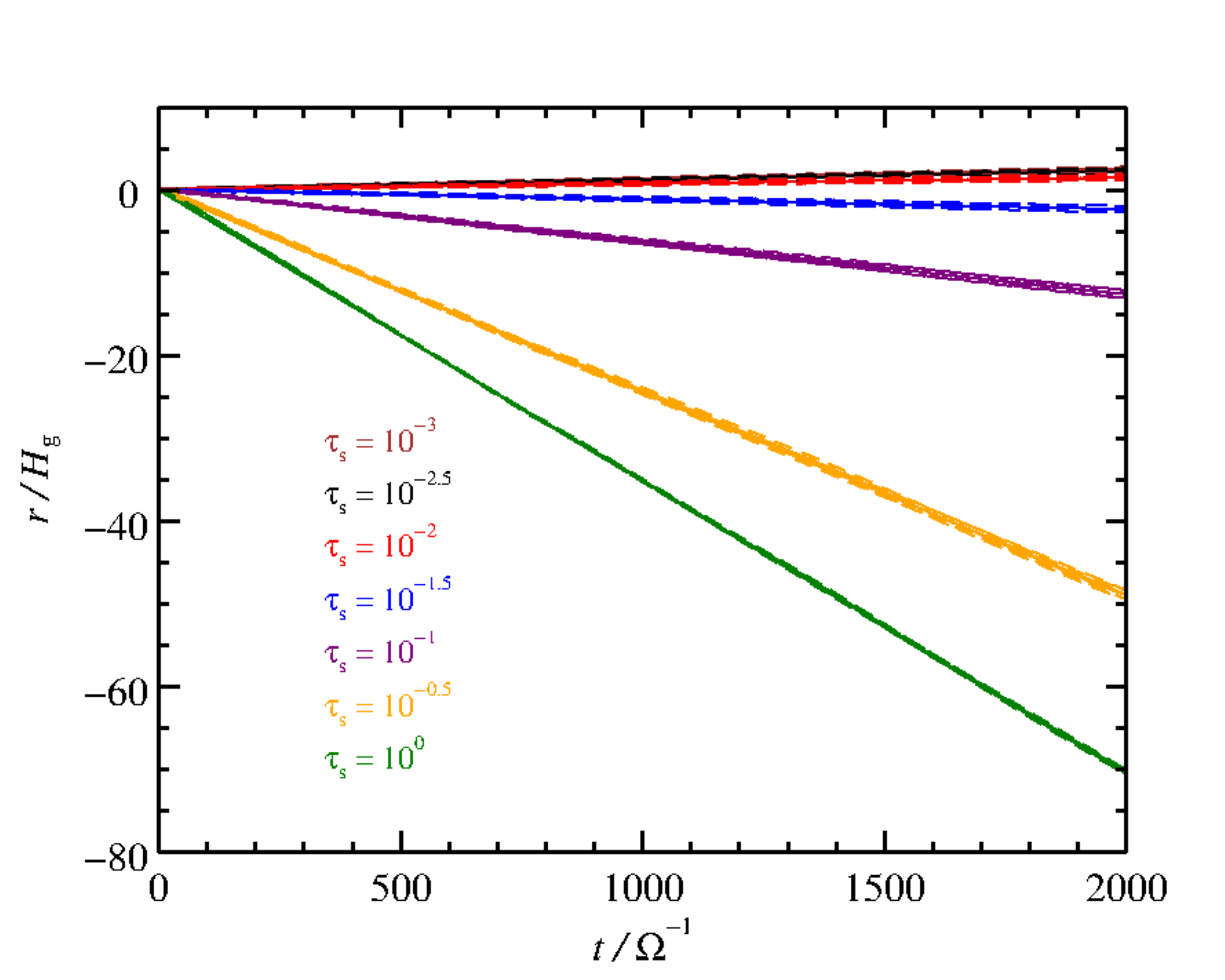}
\caption{Evolution of particle radial positions with respect to time in the case of run SI30-3-2-d. The smallest species drift outwards, while species with sizes $\tau_{\rm{s}} > 10^{-2}$ drift inwards. Compared to the run with continuous size distribution shown in Fig. \ref{rvst} the spreading of the positions is smaller because of the lack of differential radial drift in these discrete species. Differential radial drift is therefor completely dominant in spreading the size distribution of particles.}
\label{rvst_disc}
\end{figure}

Figure \ref{rvst} shows the radial distance traveled by the particles in run SI30-3-2-c. In this run, the size distribution of the particles is continuous thus we grouped them into six bins for illustration purposes. The solid lines show the mean position of the particles in each bin, while the dashed lines correspond to the 1$\sigma$ variation. Each particle size bin is labeled by different colors. We see that, as expected, different particle bins experience varying amounts of radial motion depending on their size. As discussed in Sect. \ref{SectionScaleHeight}, this is partly determined by their vertical position with respect to the midplane. Both radial drift and differential radial drift are strongest for the largest species. The position of the small species increases with time, implying outward transport in the disk. 
As the particles are distributed continuously in terms of their size, differential radial drift dominates over diffusion in spreading the distribution of particle positions. 

Figure \ref{rvst_disc} shows the evolution of the particle radial positions for run SI30-3-2-d. Particles with  $\tau_{\rm{s}} > 10^{-2}$ drift inwards, while species with $\tau_{\rm{s}} \le 10^{-2}$ move outwards. Compared to Fig. \ref{rvst}, the level of spreading for all species is significantly lower, since differential drift does not play a role. 

Overall, our results imply that differential radial drift contributes more to the spreading of particle species than diffusion. However, as evident in Fig. \ref{rvst}, differential drift does not mix the particle species. In fact, diffusion must be present to achieve mixing between particles of different sizes.

The dynamics of solids driven by the streaming instability, once the realistic case of multiple particle species is considered, may be responsible for the transport of chondrules as well as matrix material in the early Solar System. The chondrule-matrix complementarity  in carbonaceous chondrites implies that both components formed in the same region of the disk and did not drift apart during transport to the chondrite forming region \citep{Hezel2010, Jacquet2014, Johansen2015}. Our results here show that differential radial drift is generally a more powerful mechanism than diffusion for separating particle species of different sizes.

\section{Summary}
\label{SectionDiscussion}

In this paper we built on the work of \cite{Bai2010a} and modeled the dynamics of multiple particle species embedded in gas. To do this, we used the Pencil Code and performed 2D fluid-particle simulations that span the radial and vertical plane. We studied three scenarios in terms of particle size, such that the Stokes number of the solid materials in a given model was $\tau_{\rm{s}} = 10^{-4} - 10^{-1}$, $\tau_{\rm{s}} = 10^{-3} - 10^{0}$ or $\tau_{\rm{s}} = 10^{-1} - 10^0$. The size distribution of the particles was distributed according to $\rm{d}\it{N}/\rm{d}\it{a} \propto \it{a^{-q}}$, where $N$ is the total number density and $a$ is the radius of the particles. We considered $q=3$ (shallow) and $q=4$ (steep) so that our assumed size distributions envelope a variety of distributions predicted by observations and numerical simulations. On one hand, we distributed the particles into discrete size bins, in order to compare with \cite{Bai2010a}. On the other hand, we further studied a more realistic case and modeled systems where the particles were distributed continuously in terms of their size.  

We showed that the dynamics of particles in protoplanetary disks changes once we consider the realistic case of multiple solid sizes embedded in the gas. Our main results can be summarized as follows:

\begin{itemize}

\item Due to the friction from the sub-Keplerian gas, the particles experience radial drift. At the same time, the large particles trigger the streaming instability and the generated turbulence drives the radial and vertical diffusion of the solid materials. 

\item Most interestingly, in Fig. \ref{rvst} and Fig. \ref{rvst_disc} we see that small particles move outwards as also observed in \cite{Bai2010a}. This behavior is not seen in previous streaming instability simulations that contain particles of a single species e.g., \cite{Johansen2007}. As the larger species stir up the gas, the gas is accelerated locally and moves outwards, taking the strongly coupled (small) dust particles with it. At the same time, larger species move inwards but at drift velocities slower than expected from the single species NSH solution (e.g., Fig. \ref{disc:f1}).

\item From the comparison of models with different size distribution exponents ($q$), particle distribution methods (discrete or continuous) and particle sizes, we see that the sizes of the participating particles is the most important factor. The strength of turbulence appears to be independent of the steepness of the size distribution (see Fig. \ref{v_3_vs_4:f1} and Fig. \ref{v_3_vs_4:f2}). Whether the particles are distributed in a discrete or continuous fashion does not produce large differences in the measured turbulence levels either (see Fig. \ref{v_d_vs_c:f1} and Fig. \ref{v_d_vs_c:f2}).

\item As indicated by the relatively uniform level of the vertical component of the gas root-mean-square velocity away from the midplane (see Fig. \ref{GasVelocityFluct} and Fig. \ref{GasVelocityFluct41}), the turbulence generated by the large particles located close to the midplane is not limited to the midplane. The turbulence extends to larger heights and as a result the smaller particle species that reside at these heights also experience stirring generated near the midplane. As a consequence, compared to single species streaming instability simulations \citep{Carrera2015}, the measured particle scale heights in our models are larger by several factors.

\end{itemize}

The results listed above have important implications for protoplanetary disks. As shown in Fig. \ref{rvst}, in the more realistic case of continuous particle size distribution, both diffusion and differential radial drift contribute to the spreading of particles over time. Contrary to differential drift, diffusion is driven by the self-generated turbulence and is present in all our systems, independent of the particle size distribution we consider. The thickness of the dust layer in protoplanetary disks is an important factor that affects the efficiency of coagulation into dust aggregates \citep{Zsom2011, Drazkowska2016}. In the future, interaction between particle species should be taken into account in planet formation models.

\begin{acknowledgements}

We thank the anonymous referee for their comments that helped improve the manuscript.
NS was funded by the ''Bottlenecks for particle growth in turbulent aerosols'' grant from the Knut and Alice Wallenberg Foundation (2014.0048). AJ is grateful for support from the KAW Foundation (grant 2012.0150), the European Research Council (ERC Consolidator Grant 724687-PLANETESYS) and the Swedish Research Council (grant 2014-5775). AJ and CCY thank the European Research Council (ERC Starting Grant 278675-PEBBLE2PLANET). The simulations were performed on resources provided by the Swedish National Infrastructure for Computing (SNIC) at LUNARC in Lund University.

\end{acknowledgements}

\bibliographystyle{aa}
\bibliography{MultiRefs}

\nopagebreak

\begin{appendix}
\section{Gas velocity in large simulation box}
\label{Appendix1}
In Sect. \ref{SectionScaleHeight} we showed the evolution of the particle scale height for run SI30-3-8-c in a simulation box of $L_x = L_z = 0.8 H_{\rm{g}}$. Compared to the other panels in Fig. \ref{hp:7} that show the same run in smaller boxes ($L_x = L_z = 0.2 H_{\rm{g}}$ and $L_x = L_z = 0.4 H_{\rm{g}}$), the equilibrium scale height of the corresponding species do not converge. The small species are diffused to larger heights than before and the larger ones settle closer to the midplane.

\begin{figure*}[h!]
\centering
\begin{subfigure}[t]{0.49\textwidth}
\includegraphics[width=1\columnwidth]{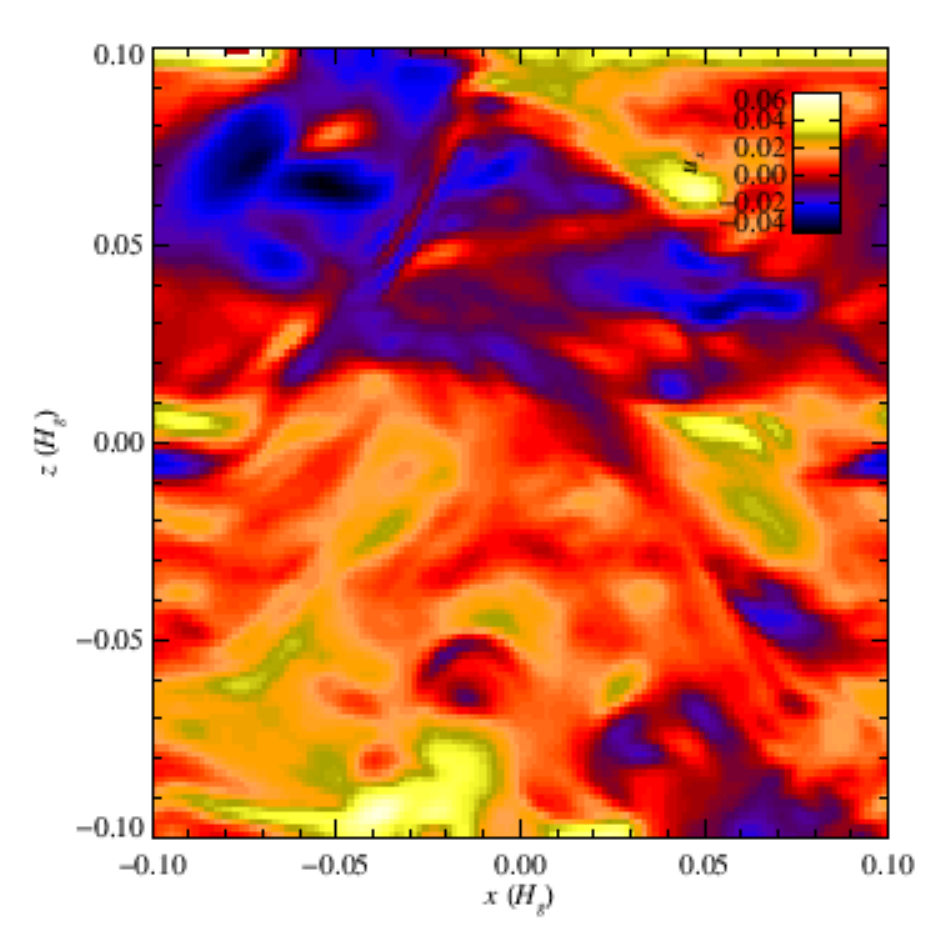}
\end{subfigure}
\begin{subfigure}[t]{0.49\textwidth}
\includegraphics[width=1\columnwidth]{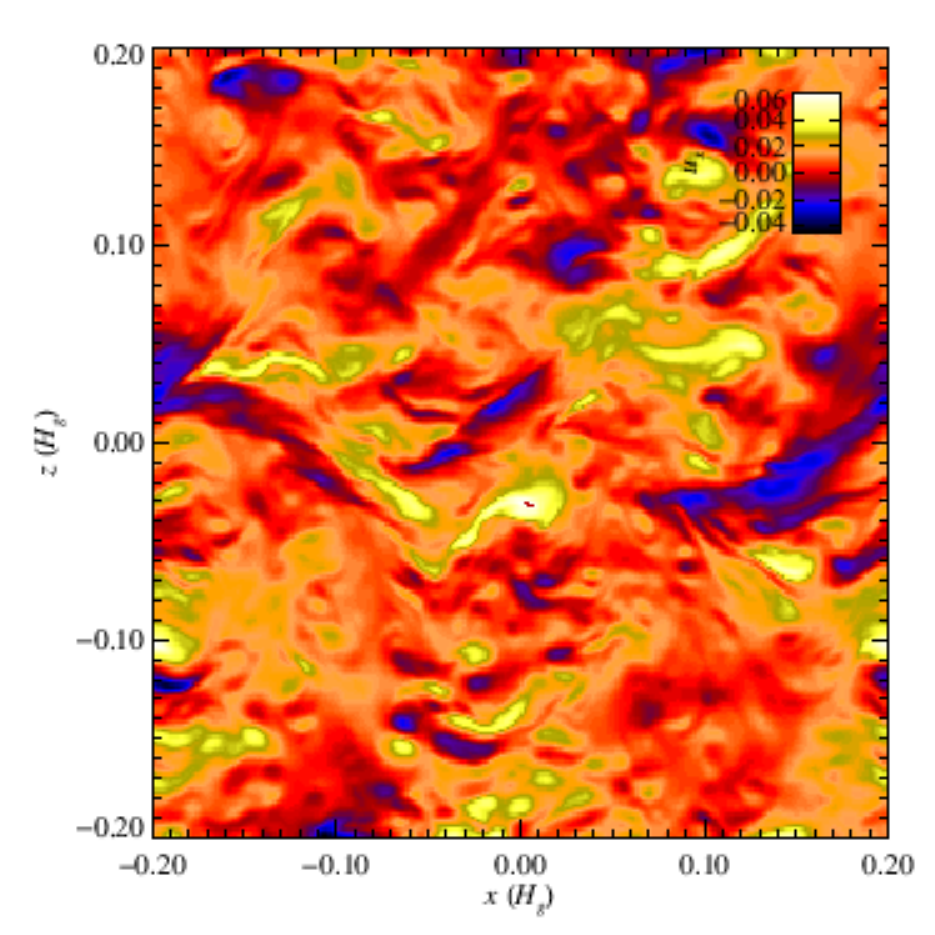}
\end{subfigure}
\begin{subfigure}[t]{0.49\textwidth}
\includegraphics[width=1\columnwidth]{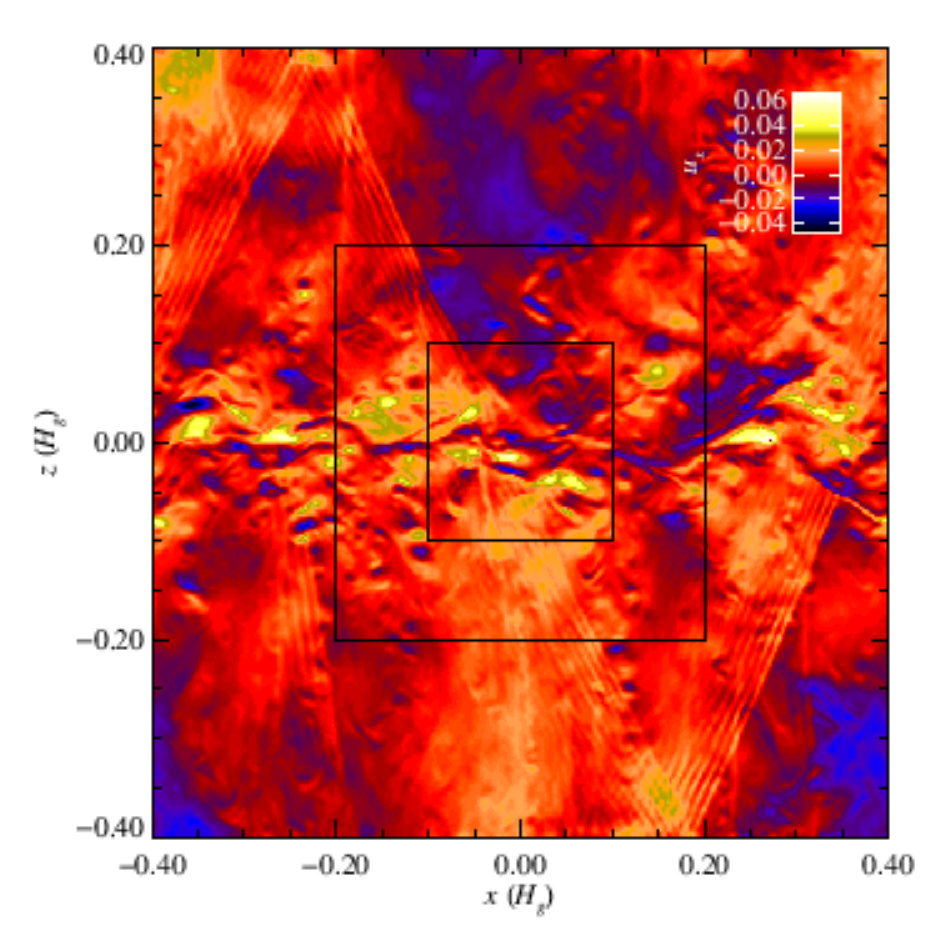}
\end{subfigure}
\caption{Radial gas velocity at $t \approx 1000 \mbox{ } \varOmega^{-1}$ corresponding to runs SI30-3-2-c, SI30-3-4-c and SI30-3-8-c, respectively. In addition to the oscillatory motion driven by the streaming instability, $u_x$ develops large-scale shock-like patterns in the largest box, shown in the final panel. Here, we overplot the outline of the two smaller box sizes for comparison.}
\label{GasPattern}
\end{figure*}

 The radial component of the gas velocity may shed some light on the discrepancy. After about 60 orbits, $u_x$ develops large-scale, diagonal shock-like patterns that persist throughout the rest of the run. In Fig. \ref{GasPattern}, we show the radial gas velocity in all three box sizes. The first two panels, which correspond to the smaller boxes, show the expected oscillatory motion due to stirring from the larger particles. The last panel, in addition to oscillatory motion, also displays shock-like behavior. This behavior was not seen in the 3D single-species simulations with large boxes in \cite{Yang2014} and is likely an artifact of the numerical prescription. As recommended in \cite{Li2018}, outflow instead of reflecting boundary conditions in the vertical direction might eliminate the shock-like patterns.

\end{appendix}

\end{document}